\documentclass[12pt,english]{article}
\usepackage{times}
\usepackage[T1]{fontenc}
\usepackage{geometry}
\usepackage{rotating}
\usepackage{graphicx}
\usepackage{setspace}
\usepackage{amssymb}

\makeatletter

\providecommand{\tabularnewline}{\\}

\usepackage{bm}

\usepackage{babel}
\makeatother
\begin{document}

\section*{\noindent On the Improvement of the Performance of Inexpensive Electromagnetic
Skins by means of an Inverse Source Design Approach}

\noindent ~

\noindent \vfill

\noindent G. Oliveri,$^{(1)(2)}$ \emph{Fellow}, \emph{IEEE}, F. Zardi,$^{(1)(2)}$
and A. Massa,$^{(1)(2)(3)(4)(5)}$ \emph{Fellow, IEEE}

\noindent \vfill

\noindent {\footnotesize ~}{\footnotesize \par}

\noindent {\footnotesize $^{(1)}$} \emph{\footnotesize ELEDIA Research
Center} {\footnotesize (}\emph{\footnotesize ELEDIA}{\footnotesize @}\emph{\footnotesize UniTN}
{\footnotesize - University of Trento)}{\footnotesize \par}

\noindent {\footnotesize DICAM - Department of Civil, Environmental,
and Mechanical Engineering}{\footnotesize \par}

\noindent {\footnotesize Via Mesiano 77, 38123 Trento - Italy}{\footnotesize \par}

\noindent \textit{\emph{\footnotesize E-mail:}} {\footnotesize \{}\emph{\footnotesize giacomo.oliveri,
francesco.zardi, andrea.massa}{\footnotesize \}@}\emph{\footnotesize unitn.it}{\footnotesize \par}

\noindent {\footnotesize Website:} \emph{\footnotesize www.eledia.org/eledia-unitn}{\footnotesize \par}

\noindent {\footnotesize $^{(2)}$} \emph{\footnotesize CNIT - \char`\"{}University
of Trento\char`\"{} ELEDIA Research Unit }{\footnotesize \par}

\noindent {\footnotesize Via Sommarive 9, 38123 Trento - Italy}{\footnotesize \par}

\noindent {\footnotesize Website:} \emph{\footnotesize www.eledia.org/eledia-unitn}{\footnotesize \par}

\noindent {\footnotesize $^{(3)}$} \emph{\footnotesize ELEDIA Research
Center} {\footnotesize (}\emph{\footnotesize ELEDIA}{\footnotesize @}\emph{\footnotesize UESTC}
{\footnotesize - UESTC)}{\footnotesize \par}

\noindent {\footnotesize School of Electronic Science and Engineering,
Chengdu 611731 - China}{\footnotesize \par}

\noindent \textit{\emph{\footnotesize E-mail:}} \emph{\footnotesize andrea.massa@uestc.edu.cn}{\footnotesize \par}

\noindent {\footnotesize Website:} \emph{\footnotesize www.eledia.org/eledia}{\footnotesize -}\emph{\footnotesize uestc}{\footnotesize \par}

\noindent {\footnotesize $^{(4)}$} \emph{\footnotesize ELEDIA Research
Center} {\footnotesize (}\emph{\footnotesize ELEDIA@TSINGHUA} {\footnotesize -
Tsinghua University)}{\footnotesize \par}

\noindent {\footnotesize 30 Shuangqing Rd, 100084 Haidian, Beijing
- China}{\footnotesize \par}

\noindent {\footnotesize E-mail:} \emph{\footnotesize andrea.massa@tsinghua.edu.cn}{\footnotesize \par}

\noindent {\footnotesize Website:} \emph{\footnotesize www.eledia.org/eledia-tsinghua}{\footnotesize \par}

\noindent {\small $^{(5)}$} {\footnotesize School of Electrical Engineering}{\footnotesize \par}

\noindent {\footnotesize Tel Aviv University, Tel Aviv 69978 - Israel}{\footnotesize \par}

\noindent \textit{\emph{\footnotesize E-mail:}} \emph{\footnotesize andrea.massa@eng.tau.ac.il}{\footnotesize \par}

\noindent {\footnotesize Website:} \emph{\footnotesize https://engineering.tau.ac.il/}{\footnotesize \par}

\noindent \vfill

\noindent \emph{This work has been submitted to the IEEE for possible
publication. Copyright may be transferred without notice, after which
this version may no longer be accessible.}

\noindent \vfill

\newpage
\section*{On the Improvement of the Performance of Inexpensive Electromagnetic
Skins by means of an Inverse Source Design Approach}

~

~

~

\begin{flushleft}G. Oliveri, F. Zardi, and A. Massa\end{flushleft}

\vfill

\begin{abstract}
\noindent A new methodology for the improvement of the performance
of inexpensive static passive electromagnetic skins (\emph{SP-EMS}s)
is presented. The proposed approach leverages on the non-uniqueness
of the inverse source problem associated to the \emph{SP-EMS} design
by decomposing the induced surface current into pre-image (\emph{PI})
and null-space (\emph{NS}) components. Successively, the unknown \emph{EMS}
layout and \emph{NS} expansion coefficients are determined by means
of an alternate minimization of a suitable cost function. This latter
quantifies the mismatch between the ideal surface current, which radiates
the user-defined target field, and that actually induced on the \emph{EMS}
layout. Results from a representative set of numerical experiments,
concerned with the design of \emph{EMS}s reflecting pencil-beam as
well as contoured target patterns, are reported to assess the feasibility
and the effectiveness of the proposed method in improving the performance
of inexpensive \emph{EMS} realizations. The measurements on an \emph{EMS}
prototype, featuring a conductive ink pattern printed on a standard
paper substrate, are also shown to prove the reliability of the synthesis
process.

\vfill
\end{abstract}
\noindent \textbf{Key words}: Static Passive \emph{EM} Skins; Smart
Electromagnetic Environment; Next-Generation Communications; Metamaterials;
Metasurfaces; Inverse Scattering; Non-Radiating Currents; Inverse
Source Formulation.

\newpage
\section{Introduction and Motivation\label{sec:Introduction}}

\noindent The development of scalable and effective technologies for
the implementation of the Smart Electromagnetic Environment (\emph{SEME})
is a research area of growing relevance in next generation communication
systems \cite{Massa 2021}-\cite{Huang 2019}. As a matter of fact,
the possibility to tailor the electromagnetic propagation according
to the wireless communication needs thanks to the \emph{SEME} solutions
has revealed potential significant improvements of the quality-of-service,
the coverage, and the data throughput of current wireless systems
\cite{Massa 2021}\cite{Barbuto 2022}-\cite{Huang 2019}. In such
a framework, static passive electromagnetic skins (\emph{SP-EMS}s)
have emerged as one of the most promising technological solutions
thanks to the minimum costs and the maximum scalability \cite{Massa 2021}\cite{Barbuto 2022}\cite{Oliveri 2021c}-\cite{Vaquero 2024}.

\noindent An \emph{SP-EMS} consists of a patterned artificial surface
\cite{Yang 2019}\cite{Alu 2024} that yields advanced field manipulation
features by properly exploiting the geometrical/physical variations
of its sub-wavelength meta-atoms \cite{Oliveri 2021c}-\cite{Oliveri 2023b}.
By avoiding active/reconfigurable components and recurring to fabrication
processes borrowed from traditional printed circuit board (\emph{PCB})
technologies, a \emph{SP-EMS} is typically less expensive than a reconfigurable
intelligent surface (\emph{RIS}), a smart repeater (\emph{SR}), or
an integrated access and backhaul (\emph{IAB}) node \cite{Yang 2022}\cite{Oliveri 2021c}-\cite{Oliveri 2023b}.
Besides the cost, both the seamless installation and the absence of
any power supply as well as the virtual transparency to wireless network
operations have motivated a strong boost in the development, the demonstration,
and the deployment of \emph{SP-EMS}s \cite{Yang 2022}\cite{Oliveri 2021c}-\cite{Oliveri 2023b}.

\noindent However, several challenges have still to be addressed for
enabling a mass production \cite{Yang 2022} and a large scale deployment
of \emph{SP-EMS}s. Indeed, the implementation of an inexpensive \emph{SP-EMS}
tens of meters wide requires much cheaper substrates than those currently
adopted in \emph{PCB} manufacturing. Moreover, single-layer meta-atoms
with thin substrates would be much preferable owing to the smaller
costs, an overall reduced weight, and a lower installation complexity\@.
On the other hand, it is known from surface electromagnetics theory
\cite{Yang 2019} that a single-layer meta-atom based on a low-quality
substrate would typically result in a poor phase linearity and a sharp
loss increase for some meta-atom configurations {[}e.g., see the design
in Fig. 2(\emph{b}){]}. Thus, the arising \emph{SP-EMS} would exhibit
a mediocre power efficiency and a weak robustness to fabrication tolerances
\cite{Yang 2019}. For this reason, current \emph{SP-EMS} prototypes
often feature standard \emph{PCB} materials with non-negligible \emph{per-meter}
costs \cite{Oliveri 2021c}-\cite{Vaquero 2024}. Therefore, there
is a great interest in demonstrating very inexpensive \emph{SP-EMS}s
still reaching a good efficiency.

\noindent To overcome, on the one hand, the issue of the costs associated
to the use of standard \emph{PCB} manufacturing in \emph{SP-EMS} engineering,
but also to avoid recurring to inexpensive \emph{EMS}s with limited/poor
performance, the inverse source (\emph{IS})-driven approach, which
has recently-introduced in \cite{Oliveri 2023} to synthesize \emph{SP-EMS}s
complying with user-defined constraints, may be taken into account.
More specifically, it has been proven that the non-uniqueness of the
\emph{EMS} design problem can be successfully exploited to synthesize
multiple and equivalent (in terms of reflected footprint pattern)
layouts so that (at least) one of those may fit user-defined requirements
and goals \cite{Oliveri 2023}. In principle, such an approach is
suitable for improving the efficiency of \emph{SP-EMS}s based on inexpensive
unit cells, but such an extension would require solving non-trivial
issues such as how to (\emph{i}) code the inexpensiveness constraint,
(\emph{ii}) apply and scale the generalized \emph{IS}-method to wide
\emph{EMS}s, and (\emph{iii}) yield inexpensive \emph{SP-EMS}s still
fulfilling the reflection requirements satisfied with their \emph{PCB}-manufactured
counterparts.

\noindent Therefore, a new approach is proposed hereinafter to improve
the wave manipulation performance of \emph{SP-EMS}s featuring inexpensive
meta-atoms. By still leveraging on the non-uniqueness of the \emph{IS}
problem associated to the \emph{SP-EMS} design, the induced surface
current is first decomposed into pre-image (\emph{PI}) and null-space
(\emph{NS}) components. Successively, the unknown \emph{EMS} layout
and \emph{NS} expansion coefficients are determined by means of an
alternate minimization of a suitable cost function. This latter quantifies
the mismatch between the ideal surface current, which radiates the
user-defined target field, and that induced on the \emph{EMS} layout.

\noindent The main innovative contributions of this work with respect
to the state of the art literature can be summarized in the following
items:

\begin{itemize}
\item the generalization of the theoretical framework presented in \cite{Oliveri 2023}
to the improvement of the performance of inexpensive \emph{SP-EMS}s
and, more in general, to the exploitation of the \emph{NS} surface
currents to synthesize \emph{SP-EMS}s fulfilling user/application-driven
requirements/constraints, but still with advanced field manipulation
properties;
\item the numerical assessment and the experimental proof of the feasibility
of inexpensive \emph{SP-EMS}s, not based on the traditional \emph{PCB}
technology and materials, affording user-defined footprint patterns
that are typical of \emph{SEME}s and (in general) next generation
wireless communication scenarios.
\end{itemize}
\noindent The outline of the paper is as follows. After the formulation
of the \emph{EMS} synthesis problem at hand (Sect. \ref{sec:Problem-Formulation}),
the \emph{IS} \emph{NS}-based method for the optimized design of inexpensive
\emph{SP-EMS}s is detailed in Sect. \ref{sec:Design}. Section \ref{sec:Results}
reports the results of a representative set of numerical experiments
to assess, also through an experimental validation, the effectiveness
and the reliability of the proposed \emph{EMS} synthesis method as
well as the improved performance of the arising \emph{SP-EMS} layouts.
Finally, some conclusions and remarks are drawn (Sect. \ref{sec:Conclusions-and-Remarks}).

\section{\noindent Design Problem Formulation\label{sec:Problem-Formulation} }

\noindent Let us consider the benchmark scenario in Fig. 1 where a
planar \emph{SP-EMS} located in the ($x$, $y$) plane is illuminated
by an incident plane wave impinging from the angular direction $\left(\theta^{inc},\varphi^{inc}\right)$
whose associated electric field is \cite{Lindell 2019}\cite{Osipov 2017}\begin{equation}
\mathbf{E}^{inc}\left(\mathbf{r}\right)\triangleq\left(E_{TE}^{inc}\widehat{\mathbf{e}}_{TE}+E_{TM}^{inc}\widehat{\mathbf{e}}_{TM}\right)\exp\left(-j\mathbf{k}^{inc}\cdot\mathbf{r}\right)\label{eq:incident wave}\end{equation}
where $\mathbf{k}^{inc}$ is the incident wave vector\begin{equation}
\begin{array}{r}
\mathbf{k}^{inc}\triangleq-k_{0}\left[\sin\left(\theta^{inc}\right)\cos\left(\varphi^{inc}\right)\widehat{\mathbf{x}}+\right.\\
\left.+\sin\left(\theta^{inc}\right)\sin\left(\varphi^{inc}\right)\widehat{\mathbf{y}}+\cos\left(\theta^{inc}\right)\widehat{\mathbf{z}}\right],\end{array}\label{eq:incident wave vector}\end{equation}
while $\mathbf{r}$ is the \emph{EMS} local position vector {[}$\mathbf{r}=\left(x,y,z\right)${]},
$k_{0}$ and $\zeta_{0}$ being the free-space wavenumber and the
intrinsic impedance, respectively. Moreover, $\widehat{\mathbf{e}}_{TE}=\frac{\mathbf{k}^{inc}\times\widehat{\mathbf{z}}}{\left|\mathbf{k}^{inc}\times\widehat{\mathbf{z}}\right|}$
and $\widehat{\mathbf{e}}_{TM}=\frac{\widehat{\mathbf{e}}_{TE}\times\mathbf{k}^{inc}}{\left|\widehat{\mathbf{e}}_{TE}\times\mathbf{k}^{inc}\right|}$
are the \emph{TE} and the \emph{TM} mode unit vectors, respectively,
while $E_{TE}^{inc}$ and $E_{TM}^{inc}$ are the corresponding complex-valued
coefficients, $\widehat{\mathbf{z}}$ is the normal to the skin surface,
and $\left|\cdot\right|$ is the vector magnitude operator. The \emph{SP-EMS}
consists of $P\times Q$ meta-atoms centered at \{$\mathbf{r}_{pq}$
($p=1,...,P$; $q=1,...,Q$)\} and it is univocally described by the
descriptor vector $\mathcal{D}$

\noindent \begin{equation}
\mathcal{D}\triangleq\left\{ \underline{d}_{pq};p=1,...,P;\, q=1,...,Q\right\} ,\label{eq:}\end{equation}
the ($p$, $q$)-th entry of which, $\underline{d}_{pq}$ ($p=1,...,P$;
$q=1,...,Q$), is characterized by a set of $L$ descriptors (i.e.,
$\underline{d}_{pq}\triangleq\left\{ d_{pq}^{\left(l\right)};\, l=1,...,L\right\} $).

\noindent The electromagnetic interactions between the incident field
and the \emph{SP-EMS} induce on the skin support $\Omega$ a surface
electromagnetic current $\mathbf{J}\left(\mathbf{r}|\mathcal{D}\right)$
($\mathbf{J}\left(\mathbf{r}|\mathcal{D}\right)=J_{x}\left(\mathbf{r}|\mathcal{D}\right)\widehat{\mathbf{x}}+J_{y}\left(\mathbf{r}|\mathcal{D}\right)\widehat{\mathbf{y}}$)
with electric and magnetic terms (i.e., $\mathbf{J}\left(\mathbf{r}|\mathcal{D}\right)\triangleq\widehat{\mathbf{z}}\times\left[\zeta_{0}\widehat{\mathbf{z}}\times\mathbf{J}^{e}\left(\mathbf{r}'|\mathcal{D}\right)+\mathbf{J}^{m}\left(\mathbf{r}'|\mathcal{D}\right)\right]$)
that radiates in the far-field region $\Theta$ (i.e., $\mathbf{r}\in\Theta$)
a reflected electric field given by\begin{equation}
\begin{array}{r}
\begin{array}{l}
\mathbf{E}^{refl}\left(\mathbf{r}|\mathcal{D}\right)=\\
\frac{jk_{0}}{4\pi}\frac{\exp\left(-jk_{0}r\right)}{r}\int_{\Omega}\mathbf{J}\left(\mathbf{r}|\mathcal{D}\right)\exp\left(jk_{0}\widehat{\mathbf{r}}\cdot\mathbf{r}'\right)\mathrm{d}\mathbf{r}'.\end{array}\end{array}\label{eq:matching field}\end{equation}
The relation among the the electric/magnetic terms of the induced
current and the \emph{SP-EMS} unit cell descriptors $\underline{d}_{pq}$
($p=1,...,P$, $q=1,...,Q$) can be expressed according to the Love's
equivalence principle \cite{Yang 2019}\cite{Oliveri 2021c} as follows 

\noindent \begin{equation}
\left\{ \begin{array}{l}
\mathbf{J}^{e}\left(\mathbf{r}|\mathcal{D}\right)=\frac{1}{\zeta_{0}}\widehat{\mathbf{z}}\times\mathbf{k}_{inc}\times\underline{\underline{\Gamma}}\left(\mathbf{r}|\mathcal{D},\mathbf{k}^{inc}\right)\cdot\mathbf{E}^{inc}\left(\mathbf{r}\right)\\
\mathbf{J}^{m}\left(\mathbf{r}|\mathcal{D}\right)=-\widehat{\mathbf{z}}\times\underline{\underline{\Gamma}}\left(\mathbf{r}|\mathcal{D},\mathbf{k}^{inc}\right)\cdot\mathbf{E}^{inc}\left(\mathbf{r}\right)\end{array}\right.\label{eq:correnti}\end{equation}
where\begin{equation}
\underline{\underline{\Gamma}}\left(\mathbf{r}|\mathcal{D},\mathbf{k}^{inc}\right)=\left[\begin{array}{cc}
\Gamma_{TE}\left(\mathbf{r}|\mathcal{D},\mathbf{k}^{inc}\right) & 0\\
0 & \Gamma_{TM}\left(\mathbf{r}|\mathcal{D},\mathbf{k}^{inc}\right)\end{array}\right]\label{eq:}\end{equation}
 is the local complex reflection matrix of an \emph{SP-EMS} with descriptors
$\mathcal{D}$. 

\noindent Under the local periodicity assumption, it turns out that
\cite{Yang 2019}\begin{equation}
\underline{\underline{\Gamma}}\left(\mathbf{r}|\mathcal{D},\mathbf{k}^{inc}\right)\approx\sum_{p=1}^{P}\sum_{q=1}^{Q}\underline{\underline{\Gamma}}\left(\underline{d}_{pq},\mathbf{k}^{inc}\right)\Pi^{pq}\left(\mathbf{r}\right)\label{eq:electric polariz}\end{equation}

\noindent where $\Pi_{pq}\left(\mathbf{r}\right)$ is the ($p$, $q$)-th
($p=1,...,P$; $q=1,...,Q$) pixel basis function centered in $\mathbf{r}_{pq}$
($\mathbf{r}_{pq}\in\Omega$), while the local reflection coefficient
for a given meta-atom featuring $\underline{d}$ and illuminated by
an incident plane wave with an incident wave vector $\mathbf{k}^{inc}$,
$\underline{\underline{\Gamma}}\left(\underline{d},\mathbf{k}^{inc}\right)$,
can be computed with analytical, numerical, hybrid, or artificial
intelligence-based methods \cite{Yang 2019}\cite{Salucci 2018c}\cite{Oliveri 2022b}.
In this paper, a full-wave commercial \emph{SW} \cite{HFSS 2021}
has been used to build a database $\mathbb{D}$ with entries \{($\underline{d}$,
$\mathbf{k}^{inc}$), $\underline{\underline{\Gamma}}\left(\underline{d},\mathbf{k}^{inc}\right)$\}.

\noindent Within the \emph{IS} framework \cite{Oliveri 2023}, a basic
formulation of the \emph{SP-EMS} synthesis problem can be then stated
as follows

\begin{quotation}
\noindent \textbf{\emph{Standard SP-EMS IS Problem}} \textbf{(}\textbf{\emph{SISP}}\textbf{)}
- Given an incident field $\mathbf{E}^{inc}\left(\mathbf{r}\right)$
($\mathbf{r}\in\Omega$), a desired reflected field $\widetilde{\mathbf{E}}^{refl}\left(\mathbf{r}\right)$
($\mathbf{r}\in\Theta$), which corresponds to the \emph{pre-image}
surface current $\mathbf{J}_{PI}\left(\mathbf{r}\right)$ ($\mathbf{r}\in\Omega$)
fulfilling the source-version of (\ref{eq:matching field})\begin{equation}
\widetilde{\mathbf{E}}^{refl}\left(\mathbf{r}\right)=\frac{jk_{0}}{4\pi}\frac{\exp\left(-jk_{0}r\right)}{r}\int_{\Omega}\widetilde{\mathbf{J}}\left(\mathbf{r}\right)\exp\left(jk_{0}\widehat{\mathbf{r}}\cdot\mathbf{r}'\right)\mathrm{d}\mathbf{r}'\label{eq: marching field (source version)}\end{equation}

\noindent ($\mathbf{J}_{PI}\to\widetilde{\mathbf{J}}$), and a meta-atom
featuring a local reflection coefficient $\underline{\underline{\Gamma}}\left(\underline{d},\mathbf{k}^{inc}\right)$,
find the descriptors of the \emph{SP-EMS} $\mathcal{D}^{opt}$ such
that \begin{equation}
\Phi\left(\mathcal{D}\right)=\frac{\int_{\Omega}\left|\mathbf{J}\left(\mathbf{r}'|\mathcal{D}\right)-\mathbf{J}_{PI}\left(\mathbf{r}'\right)\right|^{2}\mathrm{d}\mathbf{r}'}{\int_{\Omega}\left|\mathbf{J}_{PI}\left(\mathbf{r}'\right)\right|^{2}\mathrm{d}\mathbf{r}'}\label{eq:matching currents}\end{equation}
is minimized (i.e., $\mathcal{D}^{opt}=\arg\left\{ \min_{\mathcal{D}}\left[\Phi\left(\mathcal{D}\right)\right]\right\} $).
\end{quotation}
\noindent Unfortunately, there is not guarantee that $\mathcal{D}^{opt}$
yields a good matching between induced, $\mathbf{J}\left(\mathbf{r}|\mathcal{D}^{opt}\right)$,
and pre-image, $\mathbf{J}_{PI}\left(\mathbf{r}\right)$, currents
(i.e., $\Phi\left(\mathcal{D}\right)\to0$), especially if inexpensive
and thin substrates are at hand. Fortunately, it is known that \emph{IS}
problems are ill-posed and their solutions are not unique. This is
the case of computing the surface current radiating the desired field
$\widetilde{\mathbf{E}}^{refl}\left(\mathbf{r}\right)$ ($\mathbf{r}\in\Theta$),
thus $\mathbf{J}_{PI}\left(\mathbf{r}\right)$ is just one of the
infinite set of solutions of (\ref{eq: marching field (source version)}).
Indeed, the surface current\begin{equation}
\widetilde{\mathbf{J}}\left(\mathbf{r}\right)=\mathbf{J}_{PI}\left(\mathbf{r}\right)+\mathbf{J}_{NS}\left(\mathbf{r}\right),\label{eq:decomposition}\end{equation}
where $\mathbf{J}_{NS}\left(\mathbf{r}\right)$ is the \emph{null-space}
surface current that satisfies the condition\begin{equation}
\mathbf{0}=\frac{jk_{0}}{4\pi}\frac{\exp\left(-jk_{0}r\right)}{r}\int_{\Omega}\mathbf{J}_{NS}\left(\mathbf{r}'\right)\exp\left(jk_{0}\widehat{\mathbf{r}}\cdot\mathbf{r}'\right)\mathrm{d}\mathbf{r}',\label{eq:sistema NS}\end{equation}
still radiates the target distribution $\widetilde{\mathbf{E}}^{refl}\left(\mathbf{r}\right)$
($\mathbf{r}\in\Theta$).

\noindent Accordingly, the original \textbf{}\emph{SISP} can be reformulated
as follows

\begin{quotation}
\noindent \textbf{\emph{Low-Complexity SP-EMS IS Problem}} \textbf{(}\textbf{\emph{LISP}}\textbf{)}
- Given an incident field $\mathbf{E}^{inc}\left(\mathbf{r}\right)$
($\mathbf{r}\in\Omega$), a desired reflected field $\widetilde{\mathbf{E}}^{refl}\left(\mathbf{r}\right)$
($\mathbf{r}\in\Theta$), and a meta-atom featuring a local reflection
coefficient $\underline{\underline{\Gamma}}\left(\underline{d},\mathbf{k}^{inc}\right)$,
find the descriptors of the \emph{SP-EMS} $\mathcal{D}^{opt}$ and
the most suitable null-space surface current $\mathbf{J}_{NS}^{opt}\left(\mathbf{r}\right)$
($\mathbf{r}\in\Omega$) such that\begin{equation}
\Phi\left(\mathcal{D},\,\mathbf{J}_{NS}\left(\mathbf{r}\right)\right)=\frac{\int_{\Omega}\left|\mathbf{J}\left(\mathbf{r}'|\mathcal{D}\right)-\left[\mathbf{J}_{PI}\left(\mathbf{r}'\right)+\mathbf{J}_{NS}\left(\mathbf{r}'\right)\right]\right|^{2}\mathrm{d}\mathbf{r}'}{\int_{\Omega}\left|\mathbf{J}_{PI}\left(\mathbf{r}'\right)+\mathbf{J}_{NS}\left(\mathbf{r}'\right)\right|^{2}\mathrm{d}\mathbf{r}'}\label{eq:matching currents}\end{equation}
is minimized (i.e., $\left(\mathcal{D}^{opt},\,\mathbf{J}_{NS}^{opt}\left(\mathbf{r}\right)\right)=\arg\left\{ \min_{\left(\mathcal{D},\,\mathbf{J}_{NS}\left(\mathbf{r}\right)\right)}\left[\Phi\left(\mathcal{D},\,\mathbf{J}_{NS}\left(\mathbf{r}\right)\right)\right]\right\} $).
\end{quotation}

\section{\noindent Solution Procedure\label{sec:Design}}

\noindent In order to solve the \emph{LISP}, let us first perform
the \emph{SVD} \cite{Bertero 1998} of the linear operator $\mathcal{L}$
in (\ref{eq:sistema NS}) (i.e., $\mathcal{L}:\,\widetilde{\mathbf{J}}\to\widetilde{\mathbf{E}}^{refl}$
being $\mathcal{L}\left(\cdot\right)\triangleq\int_{\Omega}\left(\cdot\right)\frac{jk_{0}\exp\left[jk_{0}\left(\widehat{\mathbf{r}}\cdot\mathbf{r}'-r\right)\right]}{4\pi r}\mathrm{d}\mathbf{r}'$)\begin{equation}
\mathcal{L}\left(\widetilde{\mathbf{J}}\right)\left(\mathbf{r}\right)=\sum_{s=1}^{S}\sigma_{s}U_{s}\left(\mathbf{r}\right)\int_{\Omega}\widetilde{\mathbf{J}}\left(\mathbf{r}'\right)V_{s}^{*}\left(\mathbf{r}'\right)\mathrm{d}\mathbf{r}'\,\,\,\,\,\mathbf{r}\in\Theta,\label{SVD}\end{equation}

\noindent where $\sigma_{s}$ is the $s$-th ($s=1,...,S$) singular
value of $\mathcal{L}\left(\cdot\right)$ with the ordering $\sigma_{s-1}$
$\ge$ $\sigma_{s}$ $\ge$ $\sigma_{s+1}$ and $\sigma_{S}\to0$
for $S\to\infty$, while $U_{s}$ and $V_{s}$ are orthogonal normalized
basis eigenfunctions in the reflected field ($U_{s}\left(\mathbf{r}\right)\in\Im\left\{ \widetilde{\mathbf{E}}^{refl}\left(\mathbf{r}\right);\,\mathbf{r}\in\Theta\right\} $)
and the surface current ($V_{s}\left(\mathbf{r}\right)\in\Im\left\{ \widetilde{\mathbf{J}}\left(\mathbf{r}\right);\,\mathbf{r}\in\Omega\right\} $)
spaces, respectively, $*$ being the adjoint operator. Therefore,
the surface current $\widetilde{\mathbf{J}}\left(\mathbf{r}\right)$
($\mathbf{r}\in\Omega$) (\ref{eq:decomposition}) can be expanded
in terms of singular values and current eigenfunctions as follows\begin{equation}
\mathbf{J}_{PI}\left(\mathbf{r}\right)=\sum_{s=1}^{s_{th}}\frac{1}{\sigma_{s}}V_{s}\left(\mathbf{r}\right)\int_{\Theta}\widetilde{\mathbf{E}}^{refl}\left(\mathbf{r}'\right)U_{s}^{*}\left(\mathbf{r}'\right)\mathrm{d}\mathbf{r}'\,\,\,\,\,\mathbf{r}\in\Omega,\label{PI}\end{equation}
\begin{equation}
\mathbf{J}_{NS}\left(\mathbf{r}|\underline{\beta}\right)=\sum_{s=s_{th}+1}^{S}\beta_{s}V_{s}\left(\mathbf{r}\right)\,\,\,\,\,\mathbf{r}\in\Omega,\label{NS}\end{equation}
where $s_{th}$ is the \emph{SVD} truncation index ($s_{th}\triangleq\arg\left[\min_{s}\left(\frac{\sigma_{s}}{\sigma_{1}}\geq\eta_{SVD}\right)\right]$,
$\eta_{SVD}$ being a user-defined threshold) and $\beta_{s}$ is
the $s$-th ($s=1,...,S$) arbitrary null-space coefficient. It is
worthwhile pointing out that $\mathcal{L}\left(\mathbf{J}_{NS}\right)\left(\mathbf{r}\right)=0$
(\ref{eq:sistema NS}) whatever the set of $\beta_{s}$ coefficients,
$\underline{\beta}=$\{$\beta_{s}$; $s>s_{th}$\}, since the set
\{$V_{s}\left(\mathbf{r}\right)$; $s>s_{th}$\} belongs to the kernel
of the operator $\mathcal{L}\left(\cdot\right)$.

\noindent Thanks to (\ref{NS}), the \emph{LISP} is then reformulated
in the following alternative manner:

\begin{quotation}
\noindent \textbf{\emph{LISP}} (\emph{SVD-Based Formulation}) - Given
an incident field $\mathbf{E}^{inc}\left(\mathbf{r}\right)$ ($\mathbf{r}\in\Omega$),
a desired reflected field $\widetilde{\mathbf{E}}^{refl}\left(\mathbf{r}\right)$
($\mathbf{r}\in\Theta$), a meta-atom featuring a local reflection
coefficient $\underline{\underline{\Gamma}}\left(\underline{d},\mathbf{k}^{inc}\right)$,
and the \emph{SVD} threshold $\eta_{SVD}$, find the descriptors of
the \emph{SP-EMS} $\mathcal{D}^{opt}$ and the most suitable set of
null-space coefficients $\underline{\beta}^{opt}$ such that\begin{equation}
\Phi\left(\mathcal{D},\,\underline{\beta}\right)=\frac{\int_{\Omega}\left|\mathbf{J}\left(\mathbf{r}"|\mathcal{D}\right)-\widetilde{\mathbf{J}}\left(\mathbf{r}"|\underline{\beta}\right)\right|^{2}\mathrm{d}\mathbf{r}"}{\int_{\Omega}\left|\widetilde{\mathbf{J}}\left(\mathbf{r}"|\underline{\beta}\right)\right|^{2}\mathrm{d}\mathbf{r}"}\label{eq:matching currents + SVD}\end{equation}
is minimized (i.e., $\left(\mathcal{D}^{opt},\,\underline{\beta}^{opt}\right)=\arg\left\{ \min_{\left(\mathcal{D},\,\underline{\beta}\right)}\left[\Phi\left(\mathcal{D},\,\underline{\beta}\right)\right]\right\} $)
being\begin{equation}
\widetilde{\mathbf{J}}\left(\mathbf{\mathbf{r}}|\underline{\beta}\right)=\sum_{s=1}^{s_{th}}\frac{1}{\sigma_{s}}V_{s}\left(\mathbf{r}\right)\int_{\Theta}\widetilde{\mathbf{E}}^{refl}\left(\mathbf{r}'\right)U_{s}^{*}\left(\mathbf{r}'\right)\mathrm{d}\mathbf{r}'+\sum_{s=s_{th}+1}^{S}\beta_{s}V_{s}\left(\mathbf{r}\right).\label{eq:Lugagna}\end{equation}

\end{quotation}
\noindent In order to minimize (\ref{eq:matching currents + SVD}),
while a simultaneous optimization of the two unknowns (i.e., $\mathcal{D}$
and $\underline{\beta}$) is in principle viable, an alternate optimization
significantly simplifies the search of $\mathcal{D}^{opt}$ and $\underline{\beta}^{opt}$
\cite{van den Berg 2002}. Such a strategy can be summarized into
the interleaving of two phases aimed at updating the sequences \{$\mathcal{D}_{n}$;
$n=1,...,N$\} and \{$\underline{\beta}_{n}$; $n=1,...,N$\} towards
$\mathcal{D}_{n}\to\mathcal{D}^{opt}$ and $\underline{\beta}_{n}\to\underline{\beta}^{opt}$,
$n$ ($n=1,...,N$) being the updating index. 

\noindent The former ({}``\emph{SP-EMS Update}'') generates the
$n$-th ($n=1,...,N$) \emph{SP-EMS} trial layout $\mathcal{D}_{n}$
by minimizing the cost function $\Phi_{\beta}\left(\mathcal{D}\right)$
given by $\Phi_{\beta}\left(\mathcal{D}\right)\triangleq\Phi\left(\mathcal{D},\,\underline{\beta}_{n-1}\right)$
(i.e., $\mathcal{D}_{n}$ $=$ $\arg$ $\left\{ \min_{\mathcal{D}}\left[\Phi\left(\mathcal{D},\,\underline{\beta}_{n-1}\right)\right]\right\} $)
being $\underline{\beta}_{0}=\underline{0}$) by means of an exhaustive
search within the off-line built database $\mathbb{D}$. More in detail,
for each ($p$, $q$)-th ($p=1,...,P$; $q=1,...,Q$) meta-atom of
the \emph{EMS} - given the incident field at hand ($\to$ $\mathbf{k}_{inc}$)
- the exhaustive process identifies in $\mathbb{D}$ the value $\left.\underline{d}_{pq}\right\rfloor _{n}$
(i.e., the most suitable entry \{($\left.\underline{d}_{pq}\right\rfloor _{n}$,
$\mathbf{k}^{inc}$), $\underline{\underline{\Gamma}}\left(\left.\underline{d}_{pq}\right\rfloor _{n},\mathbf{k}^{inc}\right)$\})
for which $\mathbf{J}\left(\mathbf{r}_{pq}|\left.\underline{d}_{pq}\right\rfloor _{n}\right)\approx\widetilde{\mathbf{J}}\left(\mathbf{r}_{pq}|\underline{\beta}_{n-1}\right)$
being $\mathbf{J}\left(\mathbf{r}_{pq}|\underline{d}_{pq}\right)$
$=$ $\widehat{\mathbf{z}}$ $\times$ $\left[\zeta_{0}\widehat{\mathbf{z}}\right.$
$\times$ $\frac{1}{\eta}$ $\widehat{\mathbf{z}}$ $\times$ $\mathbf{k}_{inc}$
$\times$ $\underline{\underline{\Gamma}}\left(\underline{d}_{pq},\mathbf{k}^{inc}\right)$
$\Pi^{pq}\left(\mathbf{r}\right)$ $\cdot$ $\mathbf{E}^{inc}\left(\mathbf{r}\right)$
$-$ $\widehat{\mathbf{z}}$ $\times$ $\underline{\underline{\Gamma}}\left(\underline{d}_{pq},\mathbf{k}^{inc}\right)$
$\Pi^{pq}\left(\mathbf{r}\right)$ $\cdot$ $\left.\mathbf{E}^{inc}\left(\mathbf{r}\right)\right]$
(Sect. \ref{sec:Problem-Formulation}). 

\noindent The second phase ({}``\emph{NS Current} \emph{Update}'')
yields the $n$-th ($n=1,...,N$) set of null-space coefficients $\underline{\beta}_{n}$
by minimizing the cost function $\Phi_{\mathcal{D}}\left(\underline{\beta}\right)$
given by $\Phi_{\mathcal{D}}\left(\underline{\beta}\right)\triangleq\Phi\left(\mathcal{D}_{n},\,\underline{\beta}\right)$
(i.e., $\underline{\beta}_{n}=\arg\left\{ \min_{\underline{\beta}}\left[\Phi\left(\mathcal{D}_{n},\,\underline{\beta}\right)\right]\right\} $)
with a multi-agent evolutionary optimization technique based on the
particle swarm mechanisms \cite{Rocca 2009}. The nested loop is terminated
if either $n=N$ or $\Phi_{\beta}\left(\mathcal{D}_{n}\right)\le\eta_{\Phi}$
or $\Phi_{\mathcal{D}}\left(\underline{\beta}_{n}\right)\le\eta_{\Phi}$
($\eta_{\Phi}$ being the convergence threshold) by outputting the
optimal setups $\mathcal{D}^{opt}=\mathcal{D}_{n}$ and $\underline{\beta}^{opt}=\underline{\beta}_{n}$.

\section{\noindent Numerical Results and Experimental Validation\label{sec:Results}}

\noindent This section is aimed at assessing the effectiveness of
the \emph{SP-EMS} design strategy presented in Sect. \ref{sec:Design}
for solving the problem formulated in Sect. \ref{sec:Problem-Formulation}
with a selected set of numerical results (Sect. \ref{sec:Results (numerical)})
and an experimental validation (Sect. \ref{sec:Results (experimental)}). 

\noindent Without loss of generality, the following benchmark scenario
has been assumed. The primary far-field source has been modeled with
a plane wave at $f=5.5$ {[}GHz{]} impinging on the \emph{SP-EMS}
from the broadside direction (i.e., $\theta^{inc}=\varphi^{inc}=0$
{[}deg{]}) with a linearly polarized \emph{TE} field of unitary magnitude
(i.e., $E_{TE}^{inc}=1$ {[}V/m{]} and $E_{TM}^{inc}=0$ {[}V/m{]})
and incident wave vector $\mathbf{k}^{inc}=-k_{0}\widehat{\mathbf{z}}$.
The control parameters in Sect. \ref{sec:Design} have been set to
$\eta_{SVD}=10^{-1}$, $\eta_{\Phi}=10^{-4}$, and $N=10^{4}$. Moreover,
the field pattern reflected from the \emph{SP-EMS} \emph{$\mathcal{D}$}
(i.e., $\mathbf{E}^{refl}\left(\mathbf{r}|\mathcal{D}\right)$, $\mathbf{r}\in\Theta$)
has been simulated with the commercial software Ansys HFSS \cite{HFSS 2021}
in a set of $M$ sampling points {[}$\to$ $S=\min\left(P\times Q,\, M\right)${]}.

\noindent A basic unit cell geometry $\Delta$-sided featuring a conductive-ink
square patch {[}i.e., $L=1$ - Fig. 2(\emph{a}){]} printed on single-layer
\emph{paper} \emph{substrate} \cite{Yang 2007a}\cite{Kawahara 2014}
has been chosen as a representative example of a very inexpensive
meta-atom. Such a choice has been motivated by the potential reduction
of the manufacturing costs (i.e., two orders of magnitude lower than
traditional \emph{PCB} technology \cite{Kawahara 2014}) as well as
its suitability for a fast and ultra-low-cost mass production \cite{Yang 2007a}.
According to the guidelines in \cite{Yang 2007a}, the paper substrate
has been modeled with a homogeneous layer having a relative permittivity
of $\varepsilon_{r}=3.2$ and a dielectric loss tangent equal to $\tan\delta=7.7\times10^{-2}$
with thickness $\tau=2.08\times10^{-3}$ {[}m{]}. Such a setup is
not a customized one since similar values model different types of
paper/cardboard in sub-6 GHz applications \cite{Alimenti 2012}-\cite{Genovesi 2016}.

\noindent The magnitude {[}Fig. 2(\emph{b}){]} and the phase {[}Fig.
2(\emph{c}){]} of the reflection coefficient, $\Gamma_{TE}\left(d^{\left(1\right)}\right)$,
as a function of the patch side $d^{\left(1\right)}$ of the meta-atom
in Fig. 2(\emph{a}) have been numerically simulated by assuming a
$0.5\,\lambda$ wide unit cell and a printing precision of $10^{-4}$
{[}m{]}, which faithfully models a typical office inkjet printer with
a resolution of $1200$ dots-per-inch. As expected, owing to the high-loss
substrate material besides the single-layer nature of the meta-atom,
the magnitude of the reflection coefficient is poor at some values
of $d^{\left(1\right)}$. For instance, $\left|\Gamma_{TE}\left(d^{\left(1\right)}\right)\right|_{d^{\left(1\right)}\approx1.4\times10^{-3}\,[m]}\approx-23.5$
{[}dB{]} {[}Fig. 2(\emph{b}){]}. Such a value is much worse than that
of typical \emph{PCB}-based meta-atoms even though printed on relatively
inexpensive substrates \cite{Yang 2019}\cite{Oliveri 2021c}\cite{Oliveri 2022}.
For comparison purposes, let us consider the case of a meta-atom printed
on an ISOLA 370HR substrate ($\varepsilon_{r}=3.92$, $\tan\delta=2.5\times10^{-2}$,
and $\tau=7.11\times10^{-4}$ {[}m{]}), plotted with the blue line
in Figs. 2(\emph{b})-2(\emph{c}), where $\left|\Gamma_{TE}\left(d^{\left(1\right)}\right)\right|\ge-11.7$
{[}dB{]} {[}Fig. 2(\emph{b}){]}.

\subsection{\noindent Numerical Assessment \label{sec:Results (numerical)}}

\noindent The first test case deals with a $P\times Q=35\times35$
\emph{SP-EMS} centered at $h=5$ {[}m{]} over the floor {[}Fig. 1(\emph{a}){]}
within an area of $A_{\Omega}\approx0.95\times0.95$ {[}$\mathrm{m}^{2}${]}.
Such a skin has been requested to reflect in the coverage area $\Theta$
of extension $A_{\Theta}\approx0.95\times0.95$ {[}$\mathrm{m}^{2}${]}
the target field distribution $\widetilde{\mathbf{E}}^{refl}\left(\mathbf{r}\right)$
($\mathbf{r}\in\Theta$) in Fig. 3(\emph{a}), which consists of a
pencil beam focused towards the direction $\left(\theta^{refl},\varphi^{refl}\right)=\left(30,\,-45\right)$
{[}deg{]}.

\noindent According to the design procedure in Sect. \ref{sec:Design},
the synthesis process starts with the \emph{SVD} \cite{Bertero 1998}
of the linear operator $\mathcal{L}$ in (\ref{eq:sistema NS}) to
determine the $S$ singular values \{$\sigma_{s}$; $s=1,...,S$\}
and the corresponding sets of eigenfunctions, \{$U_{s}\left(\mathbf{r}\right)$
($\mathbf{r}\in\Theta$); $s=1,...,S$\} and \{$V_{s}\left(\mathbf{r}\right)$
($\mathbf{r}\in\Omega$); $s=1,...,S$\} so that the know terms in
(\ref{eq:Lugagna}) are defined and the unknowns in (\ref{eq:matching currents + SVD})
to be optimized are $\mathcal{D}$ and $\underline{\beta}$. The plot
of the normalized $\mathcal{L}$ spectrum (i.e., \{$\widehat{\sigma}_{s}$;
$s=1,...,S$\} being $\widehat{\sigma}_{s}\triangleq\frac{\sigma_{s}}{\sigma_{1}}$)
in Fig. 3(\emph{b}) exhibits the well-known {}``knee'' behaviour
and the number of singular values above the threshold $\eta_{SVD}$
turns out to be approximately $s_{th}\approx335$.

\noindent Figure 4 shows the null-space coefficients $\underline{\beta}^{opt}$
and the \emph{SP-EMS} layout $\mathcal{D}^{opt}$ synthesized at the
converge ($n=N$) of the optimization process to minimize (\ref{eq:matching currents + SVD})
(Sect. \ref{sec:Design}), while the field reflected by the \emph{SP-EMS}
in the far-field region $\Theta$\emph{,} $\mathbf{E}^{refl}\left(\mathbf{r}|\mathcal{D}\right)$
\emph{}($\mathbf{r}\in\Theta$)\emph{,} is given in Fig. 5(\emph{a}).
This latter distribution turns out to be quite close to the target
one $\widetilde{\mathbf{E}}^{refl}\left(\mathbf{r}\right)$ ($\mathbf{r}\in\Theta$)
{[}Fig. 3(\emph{a}){]} with a faithful generation of the pencil beam
along the right angular direction $\left(\theta^{refl},\varphi^{refl}\right)=\left(30,\,-45\right)$
{[}deg{]}.

\noindent In order to detail the features of the \emph{NS}-based \emph{SP-EMS}
synthesis, let us analyze the behaviour of the null-space coefficients
$\underline{\beta}^{opt}$ and related quantities {[}i.e., $\widetilde{\mathbf{J}}\left(\mathbf{\mathbf{r}}|\underline{\beta}\right)$
and $\mathbf{J}_{NS}\left(\mathbf{r}|\underline{\beta}\right)$ ($\mathbf{r}\in\Omega$){]}
together with the corresponding footprint fields {[}Figs. 5(\emph{c})-5(\emph{d}){]}.
As expected from \emph{IS} theory \cite{Oliveri 2023}, there exist
a \emph{NS} current $\mathbf{J}_{NS}\left(\mathbf{r}|\underline{\beta}\right)$
($\mathbf{r}\in\Omega$) {[}Figs. 6(\emph{c})-6(\emph{d}){]} corresponding
to the non-zero magnitude entries of $\underline{\beta}^{opt}$ {[}Fig.
4(\emph{a}){]} that, by definition, radiates in far-field the null
field $\mathbf{E}_{NS}^{refl}\left(\mathbf{r}|\underline{\beta}\right)$
\emph{}($\mathbf{r}\in\Theta$) {[}Fig. 5(\emph{c}){]} and that, when
superimposed to the \emph{PI} current term in Fig. 6(\emph{a})-6(\emph{b})
{[}i.e., $\widetilde{\mathbf{J}}\left(\mathbf{\mathbf{r}}|\underline{\beta}\right)=\mathbf{J}_{PI}\left(\mathbf{r}|\underline{\beta}\right)+\mathbf{J}_{NS}\left(\mathbf{r}|\underline{\beta}\right)$
($\mathbf{r}\in\Omega$) - Figs. 6(\emph{e})-6(\emph{f}){]}, does
not perturb the \emph{PI} footprint {[}i.e., $\mathbf{E}_{TOT}^{refl}\left(\mathbf{r}|\underline{\beta}\right)\approx\mathbf{E}_{PI}^{refl}\left(\mathbf{r}|\underline{\beta}\right)$
\emph{}($\mathbf{r}\in\Theta$) - Fig. 5(\emph{d}) vs. Fig. 5(\emph{b}){]},
this latter being a close approximation of the target one {[}i.e.,
$\mathbf{E}_{PI}^{refl}\left(\mathbf{r}|\underline{\beta}\right)\approx\widetilde{\mathbf{E}}^{refl}\left(\mathbf{r}\right)$
\emph{}($\mathbf{r}\in\Theta$) - Fig. 5(\emph{b}) vs. Fig. 3(\emph{a}){]}
owing to (\ref{PI}). Still concerning the synthesized $\underline{\beta}^{opt}$
vector, it is worth observing the highly-irregular phase profile in
Fig. 4(\emph{a}), which is somehow generally \emph{a-priori} unpredictable.
This is a further motivation for choosing in Sect. \ref{sec:Design}
a global optimizer to minimize the cost function $\Phi_{\mathcal{D}}\left(\underline{\beta}\right)$,
besides the need of facing the nonlinear nature of this latter.

\noindent Next, let us complete the discussion on the features of
the \emph{NS}-based approach by focusing on $\mathcal{D}^{opt}$ in
comparison with $\mathcal{D}^{PI}$, which are the \emph{SP-EMS} descriptors
yielded by just matching the \emph{PI} currents in (\ref{eq:matching currents + SVD})
(i.e., $\mathcal{D}^{PI}=\mathcal{D}_{1}$ where $\mathcal{D}_{1}=\arg\left\{ \min_{\mathcal{D}}\left[\Phi\left(\mathcal{D},\,\underline{\beta}_{0}\right)\right]\right\} $
being $\underline{\beta}_{0}=\underline{0}$), thus neglecting the
\emph{NS} contribution. Figure 7 shows the layout of the $\mathcal{D}^{PI}$-coded
\emph{SP-EMS} {[}Fig. 7(\emph{a}){]} and the corresponding reflected
footprint $\mathbf{E}^{refl}\left(\mathbf{r}|\mathcal{D}^{PI}\right)$
($\mathbf{r}\in\Theta$) {[}Fig. 7(\emph{b}){]}. In order to highlight
the improvement granted by the exploitation of the \emph{NS} contribution,
Figure 7(\emph{c}) shows the map of the local power improvement index
$\mathcal{P}_{\mathcal{D}'\mathcal{D}"}\left(\mathbf{r}\right)$,\begin{equation}
\mathcal{P}_{\mathcal{D}'\mathcal{D}"}\left(\mathbf{r}\right)\triangleq\frac{\left[\left|\mathbf{E}^{refl}\left(\mathbf{r}|\mathcal{D}^{'}\right)\right|^{2}-\left|\mathbf{E}^{refl}\left(\mathbf{r}|\mathcal{D}^{"}\right)\right|^{2}\right]}{\max_{\mathbf{r}\in\Theta}\left\{ \left|\mathbf{E}^{refl}\left(\mathbf{r}|\mathcal{D}^{"}\right)\right|^{2}\right\} },\label{eq:}\end{equation}
of $\mathcal{D}^{opt}$ ($\mathcal{D}'\leftarrow\mathcal{D}^{opt}$)
vs. $\mathcal{D}^{PI}$ ($\mathcal{D}"\leftarrow\mathcal{D}^{PI}$)
in the far-field region $\Theta$ ($\mathbf{r}\in\Theta$). One can
infer that there is a peak power improvement exactly along the target
direction (i.e., $\theta_{max}=\theta^{refl}$, $\theta_{max}\triangleq\arg\left\{ \max_{\mathbf{r}\in\Theta}\left[\mathcal{P}_{\mathcal{D}^{opt}\mathcal{D}^{PI}}\left(\mathbf{r}\right)\right]\right\} $),
which amounts to $\mathcal{P}_{\max}\triangleq\max_{\mathbf{r}\in\Theta}\left[\mathcal{P}_{\mathcal{D}^{opt}\mathcal{D}^{PI}}\left(\mathbf{r}\right)\right]\approx28$
\%, that has been yielded without making the \emph{SP-EMS} architecture
more complex (e.g., multi-layered) or using more expensive materials,
but just exploiting the non-uniqueness of the \emph{IS} problem at
hand.

\noindent The last study carried out on the first test case has been
devoted to assess the effectiveness of the proposed \emph{SP-EMS}
synthesis approach in overcoming the intrinsic limitations of inexpensive
substrates (e.g., here a paper substrate) to reach performance closer
to those of circuit-graded materials. Towards this end, a \emph{PI}-based
\emph{SP-EMS} has been synthesized using an ISOLA substrate {[}$\mathcal{D}=\mathcal{D}_{ISOLA}^{PI}$
- Fig. 8(\emph{a}){]} and the resulting footprint pattern $\mathbf{E}^{refl}\left(\mathbf{r}|\mathcal{D}_{ISOLA}^{PI}\right)$
($\mathbf{r}\in\Theta$) {[}Fig. 8(\emph{b}){]} has been compared
with those in Fig. 7(\emph{b}) {[}i.e., $\mathbf{E}^{refl}\left(\mathbf{r}|\mathcal{D}_{paper}^{PI}\right)$
($\mathbf{r}\in\Theta$){]} and Fig. 5(\emph{a}) {[}i.e., $\mathbf{E}^{refl}\left(\mathbf{r}|\mathcal{D}_{paper}^{opt}\right)$
($\mathbf{r}\in\Theta$){]}. More specifically, the plots of the reflected
fields in the $\varphi^{refl}=-45$ {[}deg{]} cut are reported in
Fig. 9(\emph{a}). As it can be noticed, the power focusing efficiency
of the \emph{NS}-based paper-printed layout turns out to be closer
to that from the \emph{PI}-based ISOLA one {[}e.g., $\Delta\mathbf{E}_{\mathcal{D}_{ISOLA}^{PI}\mathcal{D}_{paper}^{opt}}^{refl}\left(\theta^{refl},\varphi^{refl}\right)\approx1.8$
{[}dB{]} vs. $\Delta\mathbf{E}_{\mathcal{D}_{ISOLA}^{PI}\mathcal{D}_{paper}^{PI}}^{refl}\left(\theta^{refl},\varphi^{refl}\right)\approx2.9$
{[}dB{]} being $\Delta\mathbf{E}_{\mathcal{D}^{'}\mathcal{D}^{"}}^{refl}\left(\theta,\varphi\right)\triangleq\frac{\left|\mathbf{E}^{refl}\left(\theta,\varphi|\mathcal{D}^{'}\right)\right|}{\left|\mathbf{E}^{refl}\left(\theta,\varphi|\mathcal{D}^{"}\right)\right|}$
- see the inset in Fig. 9(\emph{a}){]}. Moreover, the $\mathcal{D}_{paper}^{opt}$-layout
improves the sidelobe control of its $PI$ counterpart (e.g., $\left.\Delta\mathbf{E}_{\mathcal{D}_{paper}^{PI}\mathcal{D}_{paper}^{opt}}^{refl}\left(\theta,\varphi^{refl}\right)\right\rfloor _{\theta=0\,\,\![deg]}\approx2.1$
{[}dB{]} and $\left.\Delta\mathbf{E}_{\mathcal{D}_{paper}^{PI}\mathcal{D}_{paper}^{opt}}^{refl}\left(\theta,\varphi^{refl}\right)\right\rfloor _{\theta=82\,\,[deg]}\approx2.98$
{[}dB{]}), while it performs analogously or better than the $\mathcal{D}_{ISOLA}^{PI}$
one (e.g., $\left.\Delta\mathbf{E}_{\mathcal{D}_{ISOLA}^{PI}\mathcal{D}_{paper}^{opt}}^{refl}\left(\theta,\varphi^{refl}\right)\right\rfloor _{\theta=0\,\,\![deg]}\approx0.7$
{[}dB{]} and $\left.\Delta\mathbf{E}_{\mathcal{D}_{ISOLA}^{PI}\mathcal{D}_{paper}^{opt}}^{refl}\left(\theta,\varphi^{refl}\right)\right\rfloor _{\theta=82\,\,\![deg]}$
$\approx$ $0.2$ {[}dB{]}). For completeness, the maps of the local
power improvement index $\mathcal{P}\left(\mathbf{r}\right)$ {[}$\mathcal{P}_{\mathcal{D}_{paper}^{PI}\mathcal{D}_{ISOLA}^{PI}}\left(\mathbf{r}\right)$
- Fig. 9(\emph{b}); $\mathcal{P}_{\mathcal{D}_{paper}^{opt}\mathcal{D}_{ISOLA}^{PI}}\left(\mathbf{r}\right)$
- Fig. 9(\emph{c}){]} are reported, as well.

\noindent The second set of numerical experiments has been aimed at
evaluating the effectiveness of the \emph{NS}-based \emph{EMS} synthesis
when varying the $\Omega$ size. More specifically, the same scenario
of the previous test case has been considered, but the number of unit
cells of the square (i.e., $P=Q$) \emph{SP-EMS} has been changed
from $P=15$ ($\to$ $A_{\Omega}\approx0.16$ {[}$\mathrm{m}^{2}${]})
up to $P=95$ ($\to$ $A_{\Omega}\approx6.7$ {[}$\mathrm{m}^{2}${]}).
The plot of $\mathcal{P}_{\max}$ versus the \emph{EMS} aperture in
Fig. 10(\emph{a}) confirms the enhancement of the maximum value of
the reflection efficiency, still along $\theta^{refl}$ since always
$\theta_{max}=\theta^{refl}$, with respect to the \emph{PI} solution
regardless of the \emph{EMS} size (i.e., $\mathcal{P}_{\max}>0$ \%).
However, one can notice that the achievable improvement is more significant
for smaller apertures (e.g., $\left.\mathcal{P}_{\max}\right\rfloor _{P=95}=3$
\% vs. $\left.\mathcal{P}_{\max}\right\rfloor _{P=35}=28$ \% vs.
$\left.\mathcal{P}_{\max}\right\rfloor _{P=15}=38$ \%). This is a
key outcome towards the implementation of wide inexpensive and high-efficiency
\emph{SP-EMS}s since it would suggest the designer to avoid monolithic
realizations, while preferring the modular ones \cite{Rocca 2022}
leveraging on small tiles for covering the \emph{EMS} aperture.

\noindent For illustrative purposes, Figure 10(\emph{b}) shows the
plots of the reflected field $\mathbf{E}^{refl}\left(\theta,\varphi^{refl}|\mathcal{D}\right)$
along the cut at $\varphi^{refl}=-45$ {[}deg{]} ($-90$ {[}deg{]}
$\le\theta\le$ $90$ {[}deg{]}) for two representative \emph{EMS}
sizes (i.e., $P=15$ and $P=55$). Quantitatively, the advantage of
exploiting a \emph{NS}-driven design in terms of power focusing efficiency
reduces from $\left.\Delta\mathbf{E}_{\mathcal{D}^{opt}\mathcal{D}^{PI}}^{refl}\left(\theta^{refl},\varphi^{refl}\right)\right\rfloor _{P=Q=15}\approx1.4$
{[}dB{]} down to $\left.\Delta\mathbf{E}_{\mathcal{D}^{opt}\mathcal{D}^{PI}}^{refl}\left(\theta^{refl},\varphi^{refl}\right)\right\rfloor _{P=Q=55}\approx0.5$
{[}dB{]}. The same behavior holds true for the sidelobe control since
at the first sidelobe position ($\theta=0$ {[}deg{]}) it turns out
that $\left.\Delta\mathbf{E}_{\mathcal{D}^{opt}\mathcal{D}^{PI}}^{refl}\left(0,\varphi^{refl}\right)\right\rfloor _{P=Q=15}\approx3.6$
{[}dB{]} vs. $\left.\Delta\mathbf{E}_{\mathcal{D}^{opt}\mathcal{D}^{PI}}^{refl}\left(0,\varphi^{refl}\right)\right\rfloor _{P=Q=55}\approx1.3$
{[}dB{]}, while at the second one ($\theta=78$ {[}deg{]} when $P=15$
and $\theta=84.4$ {[}deg{]} when $P=55$) the values are $\left.\Delta\mathbf{E}_{\mathcal{D}^{opt}\mathcal{D}^{PI}}^{refl}\left(78,\varphi^{refl}\right)\right\rfloor _{P=Q=15}\approx4.1$
{[}dB{]} vs. $\left.\Delta\mathbf{E}_{\mathcal{D}^{opt}\mathcal{D}^{PI}}^{refl}\left(84.4,\varphi^{refl}\right)\right\rfloor _{P=Q=55}\approx0.8$
{[}dB{]}.

\noindent The third numerical assessment is concerned with the dependence
of the \emph{EMS} synthesis results on the target reflection direction.
Towards this end, an analysis on a square \emph{EMS} with $P=Q=35$
atoms affording a pencil beam in the azimuth plane $\varphi^{refl}=-45$
{[}deg{]} and along the variable elevation $\theta^{refl}$ ($20$
{[}deg{]} $\le\theta^{refl}\le$ $50$ {[}deg{]}) has been carried
out. The outcomes are summarized in Fig. 11 where the dependence of
$\mathcal{P}_{\max}$ on the scan direction is shown. One can observe
that the power efficiency improvement granted by the proposed approach
is non-negligible ($\left.\mathcal{P}_{\max}\right\rfloor _{\theta^{refl}}>24.5$
{[}dB{]}) also at the wider scan angles (e.g., $\left.\mathcal{P}_{\max}\right\rfloor _{\theta^{refl}=50\,\,[deg]}\approx24.7$
{[}dB{]}), when the use of poor/inexpensive substrates becomes more
and more critical, and the maximum range of variation of $\mathcal{P}_{\max}$
amounts to $\sim3$ {[}dB{]} within the angular range under test.

\noindent The last set of numerical experiments deals with more complex
coverage tasks. Indeed, the reflected pattern has been required to
comply with a contoured footprint modeling a realistic operative scenario
({}``Gare du Nord - Paris'' - Fig. 12) instead of focusing in a
target direction like in the previous pencil beam test cases. More
in detail, a $35\times35$-cell \emph{SP-EMS} located at $h=10$ {[}m{]}
over the street floor {[}Fig. 12(\emph{a}){]} has been designed to
cover either the irregular region $\mathcal{A}_{1}$ {[}Fig. 12(\emph{b}){]}
or both regions $\mathcal{A}_{1}$ and $\mathcal{A}_{2}$ {[}Fig.
12(\emph{b}){]}.

\noindent Once again, there is a non-negligible pros in using the
\emph{EMS} synthesis based on the \emph{NS} currents as pointed out
by the maps of the local power improvement index $\mathcal{P}_{\mathcal{D}^{opt}\mathcal{D}^{PI}}\left(\mathbf{r}\right)$
($\mathbf{r}\in\Theta$) in Fig. 13(\emph{b}) and Fig. 14(\emph{b}),
respectively, where the peak of the power efficiency improvement turns
out to be close to $\mathcal{P}_{\max}\approx20$ \% (vs. $\mathcal{P}_{\max}\approx28$
\% for the pencil beam case) in both test cases despite the complex
coverage requirements. For completeness, the color level plots of
the reflected field $\mathbf{E}^{refl}\left(\mathbf{r}|\mathcal{D}^{opt}\right)$
($\mathbf{r}\in\Theta$) are reported in Fig. 13(\emph{a}) and Fig.
14(\emph{a}), as well.

\subsection{Experimental Assessment \label{sec:Results (experimental)}}

\noindent In order to experimentally assess the reliability of the
proposed \emph{EMS} synthesis method, a small-scale cardboard-printed
\emph{SP-EMS} prototype has been manufactured and measured (Fig. 15).
More in detail, the optimized $P\times Q=15\times15$ design evaluated
in Fig. 10(\emph{b}) has been fabricated by depositing a conductive
ink on a standard cardboard with thickness $2.08\times10^{-3}$ {[}m{]}.
To comply with the printing area of the available Voltera V-One printer
{[}Fig. 15(\emph{a}){]}, the monolithic \emph{EMS} panel has been
subdivided in $5\times3$ parts then assembled by adding an adhesive
copper sheet as the ground-plane. Successively, the overall arrangement
has been mounted on a wooden panel to guarantee the rigidity of the
structure when undergoing the experimental measurement phase {[}Fig.
15(\emph{b}){]}\@.

\noindent As shown in Fig. 15, the agreement between measured and
simulated values of the normalized field pattern, reflected by the
\emph{SP-EMS} when illuminated by a linearly polarized \emph{TE} field,
is very satisfactory.

\section{\noindent Conclusions\label{sec:Conclusions-and-Remarks}}

\noindent An innovative technique for the improvement of the performance
of inexpensive \emph{SP-EMS}s has been presented. By leveraging on
the non-uniqueness of the \emph{IS} problem associated to the \emph{SP-EMS}
design, the surface current induced on the \emph{EMS} aperture has
been decomposed into \emph{PI} and \emph{NS} components. Successively,
the unknown \emph{EMS} layout and \emph{NS} expansion coefficients
have been determined through an alternate minimization of the mismatch
between the ideal surface current, which radiates the user-defined
target field, and that induced on the \emph{EMS} layout. Results from
a representative set of numerical experiments, concerned with the
design of \emph{EMS}s reflecting pencil-beam as well as contoured
target patterns, have been reported to assess the feasibility and
the effectiveness of the proposed method in improving the performance
of inexpensive \emph{EMS} realizations. The measurements on an \emph{EMS}
prototype, featuring a conductive ink pattern printed on a standard
paper substrate, have been also shown to prove the reliability of
the synthesis process.

\noindent From the numerical validation and performance assessment,
the following main outcomes can be drawn: (\emph{a}) the proposed
\emph{SP-EMS} synthesis method enables a non-negligible improvement,
in terms of reflected power control, over traditional (i.e., \emph{PI}-based)
design approaches when adopting inexpensive \emph{EMS} meta-atoms;
(\emph{b}) the performance improvement is more significant for smaller
\emph{EMS} apertures, thus one can infer that it is more efficient
to implement a \emph{SP-EMS} by assembling small modular tiles instead
of realizing a wide monolithic support; (\emph{c}) the \emph{NS}-based
synthesis is competitive in dealing with simple (e.g., pencil beam)
as well as complex (e.g., shaped beam) footprint requirements, while
the performance (within a quite large range) are almost independent
on the reflection angle $\left(\theta^{refl},\varphi^{refl}\right)$;
(\emph{d}) the proposed method is reliable since it carefully predicts
the performance of \emph{EMS} prototypes (Fig. 15).

\noindent Future works, beyond the scope of this manuscript, will
be aimed at extending the previous design strategy to dynamically-adaptive
architectures such as \emph{RIS}s. Thanks to the generality of the
proposed approach, the possibility to include further design constraints
is under investigation.

\section*{\noindent Acknowledgements}

\noindent This work benefited from the networking activities carried
out within the project DICAM-EXC (Departments of Excellence 2023-2027,
grant L232/2016) funded by the Italian Ministry of Education, Universities
and Research (MUR), the Project \char`\"{}Smart ElectroMagnetic Environment
in TrentiNo - SEME@TN\char`\"{} funded by the Autonomous Province
of Trento (CUP: C63C22000720003), the Project \char`\"{}AURORA - Smart
Materials for Ubiquitous Energy Harvesting, Storage, and Delivery
in Next Generation Sustainable Environments\char`\"{} funded by the
Italian Ministry for Universities and Research within the PRIN-PNRR
2022 Program, and the following projects funded by the European Union
- NextGenerationEU within the PNRR Program: Project \char`\"{}ICSC
National Centre for HPC, Big Data and Quantum Computing (CN HPC)\char`\"{}
(CUP: E63C22000970007), Project \char`\"{}Telecommunications of the
Future (PE00000001 - program \char`\"{}RESTART\char`\"{}, Structural
Project 6GWINET)'' (CUP: D43C22003080001), Project ''INSIDE-NEXT
- Indoor Smart Illuminator for Device Energization and Next-Generation
Communications'' (CUP: E53D23000990001), and Project \char`\"{}Telecommunications
of the Future (PE00000001 - program {}``RESTART'', Focused Project
MOSS)'' (CUP: J33C22002880001). A. Massa wishes to thank E. Vico
for her never-ending inspiration, support, guidance, and help.

\section*{FIGURE CAPTIONS}

\begin{itemize}
\item \textbf{Figure 1.} \emph{Mathematical Formulation} - Sketch of the
\emph{SP-EMS} design problem.
\item \textbf{Figure 2.} \emph{Illustrative Example -} Sketch of the meta-atom
geometry (\emph{a}) and plots of (\emph{b}) the magnitude and (\emph{c})
the phase of $\Gamma_{TE}\left(d^{\left(1\right)}\right)$ vs. $d^{\left(1\right)}$.
\item \textbf{Figure 3.} \emph{Illustrative Example} ($h=5$ {[}m{]}, $P\times Q=35\times35$)
- Plots of (\emph{a}) $\left|\widetilde{\mathbf{E}}^{refl}\left(\mathbf{r}\right)\right|$
($\mathbf{r}\in\Theta$) and (\emph{b}) the normalized $\mathcal{L}$
spectrum \{$\widehat{\sigma}_{s}$; $s=1,...,S$\}.
\item \textbf{Figure 4.} \emph{Illustrative Example} ($h=5$ {[}m{]}, $P\times Q=35\times35$,
\emph{}$\left(\theta^{refl},\varphi^{refl}\right)=\left(30,-45\right)$
{[}deg{]}, paper substrate) - Plots of (\emph{a}) $\underline{\beta}^{opt}$
and (\emph{b}) the \emph{SP-EMS} $\mathcal{D}^{opt}$ layout.
\item \textbf{Figure 5.} \emph{Illustrative Example} ($h=5$ {[}m{]}, $P\times Q=35\times35$,
\emph{}$\left(\theta^{refl},\varphi^{refl}\right)=\left(30,-45\right)$
{[}deg{]}, paper substrate) - Plots of (\emph{a}) $\left|\mathbf{E}^{refl}\left(\mathbf{r}|\mathcal{D}^{opt}\right)\right|$,
(\emph{b}) $\left|\mathbf{E}_{PI}^{refl}\left(\mathbf{r}\right)\right|$,
(\emph{c}) $\left|\mathbf{E}_{NS}^{refl}\left(\mathbf{r}\right)\right|$,
and (\emph{d}) $\left|\mathbf{E}_{TOT}^{refl}\left(\mathbf{r}\right)\right|$
($\mathbf{r}\in\Theta$).
\item \textbf{Figure 6.} \emph{Illustrative Example} ($h=5$ {[}m{]}, $P\times Q=35\times35$,
\emph{}$\left(\theta^{refl},\varphi^{refl}\right)=\left(30,-45\right)$
{[}deg{]}, electric $\widehat{\mathbf{y}}$ component) - Plots of
(\emph{a})(\emph{c})(\emph{e}) the magnitude and (\emph{b})(\emph{d})(\emph{f})
the phase of (\emph{a})(\emph{b}) $\mathbf{J}_{PI}\left(\mathbf{r}\right)$,
(\emph{c})(\emph{d}) $\mathbf{J}_{NS}\left(\mathbf{r}|\underline{\beta}\right)$,
and (\emph{e})(\emph{f}) $\widetilde{\mathbf{J}}\left(\mathbf{\mathbf{r}}|\underline{\beta}\right)$
($\mathbf{r}\in\Omega$).
\item \textbf{Figure 7.} \emph{Illustrative Example} ($h=5$ {[}m{]}, $P\times Q=35\times35$,
$\left(\theta^{refl},\varphi^{refl}\right)=\left(30,-45\right)$ {[}deg{]},
paper substrate) - Pictures of (\emph{a}) the \emph{SP-EMS} $\mathcal{D}^{PI}$
layout (\emph{a}) and maps of the corresponding (\emph{b}) $\left|\mathbf{E}^{refl}\left(\mathbf{r}|\mathcal{D}^{PI}\right)\right|$
and (\emph{c}) $\mathcal{P}_{\mathcal{D}^{opt}\mathcal{D}^{PI}}\left(\mathbf{r}\right)$
distributions ($\mathbf{r}\in\Theta$).
\item \textbf{Figure 8.} \emph{Illustrative Example} ($h=5$ {[}m{]}, $P\times Q=35\times35$,
$\left(\theta^{refl},\varphi^{refl}\right)=\left(30,-45\right)$ {[}deg{]},
ISOLA substrate) - Pictures of (\emph{a}) the \emph{SP-EMS} $\mathcal{D}_{ISOLA}^{PI}$
layout and map of the corresponding (\emph{b}) $\left|\mathbf{E}^{refl}\left(\mathbf{r}|\mathcal{D}^{PI}\right)\right|$
distribution ($\mathbf{r}\in\Theta$).
\item \textbf{Figure 9.} \emph{Illustrative Example} ($h=5$ {[}m{]}, $P\times Q=35\times35$,
$\left(\theta^{refl},\varphi^{refl}\right)=\left(30,-45\right)$ {[}deg{]})
- Plots of (\emph{a}) $\left|\mathbf{E}^{refl}\left(\mathbf{r}|\mathcal{D}\right)\right|$
in the $\varphi=\varphi^{refl}$-cut and maps of (\emph{b}) $\mathcal{P}_{\mathcal{D}_{paper}^{PI}\mathcal{D}_{ISOLA}^{PI}}\left(\mathbf{r}\right)$
and (\emph{c}) $\mathcal{P}_{\mathcal{D}_{paper}^{opt}\mathcal{D}_{isola}^{PI}}\left(\mathbf{r}\right)$
distributions ($\mathbf{r}\in\Theta$).
\item \textbf{Figure 10.} \emph{Numerical Results} ($h=5$ {[}m{]}, $\left(\theta^{refl},\varphi^{refl}\right)=\left(30,-45\right)$
{[}deg{]}) - Plots of (\emph{a}) $\mathcal{P}_{\max}$ versus $P$
($=Q$) and (\emph{b}) $\left|\mathbf{E}^{refl}\left(\mathbf{r}|\mathcal{D}\right)\right|$
in the $\varphi=\varphi^{refl}$-cut when $P\times Q=\left\{ 15\times15,95\times95\right\} $.
\item \textbf{Figure 11.} \emph{Numerical Results} ($h=5$ {[}m{]}, $P\times Q=35\times35$,
$\varphi^{refl}=-45$ {[}deg{]}) - Plots of $\mathcal{P}_{\max}$
versus the reflection angle $\theta^{refl}$.
\item \textbf{Figure 12.} \emph{Numerical Results} ($h=10$ {[}m{]}, $P\times Q=35\times35$;
Gare du Nord - Paris) - Visualization of (\emph{a}) the 3D view and
(\emph{b}) the aerial perspective of the \emph{EM} scenario.
\item \textbf{Figure 13.} \emph{Numerical Results} ($h=10$ {[}m{]}, $P\times Q=35\times35$;
Gare du Nord - Paris, $\mathcal{A}_{1}$ Coverage) - Maps of (\emph{a})
$\left|\mathbf{E}^{refl}\left(\mathbf{r}|\mathcal{D}^{opt}\right)\right|$
and (\emph{b}) $\mathcal{P}_{\mathcal{D}^{opt}\mathcal{D}^{PI}}\left(\mathbf{r}\right)$
($\mathbf{r}\in\Theta$).
\item \textbf{Figure 14.} \emph{Numerical Results} ($h=10$ {[}m{]}, $P\times Q=35\times35$;
Gare du Nord - Paris, $\mathcal{A}_{1}\cup\mathcal{A}_{2}$ Coverage)
- Maps of (\emph{a}) $\left|\mathbf{E}^{refl}\left(\mathbf{r}|\mathcal{D}^{opt}\right)\right|$
and (\emph{b}) $\mathcal{P}_{\mathcal{D}^{opt}\mathcal{D}^{PI}}\left(\mathbf{r}\right)$
($\mathbf{r}\in\Theta$).
\item \textbf{Figure 15.} \emph{Experimental Results} ($h=5$ {[}m{]}, $\left(\theta^{refl},\varphi^{refl}\right)=\left(30,-45\right)$
{[}deg{]}, $P\times Q=15\times15$, $\mathcal{D}_{paper}^{opt}$)
- Pictures of (\emph{a}) the fabrication process, (\emph{b}) the \emph{SP-EMS}
prototype, and (\emph{c}) the plot of $\left|\mathbf{E}^{refl}\left(\mathbf{r}|\mathcal{D}\right)\right|$
in the $\varphi=\varphi^{refl}$-cut.
\end{itemize}
~

\newpage
\begin{center}~\vfill\end{center}

\begin{center}\begin{tabular}{c}
\includegraphics[%
  width=0.95\columnwidth,
  keepaspectratio]{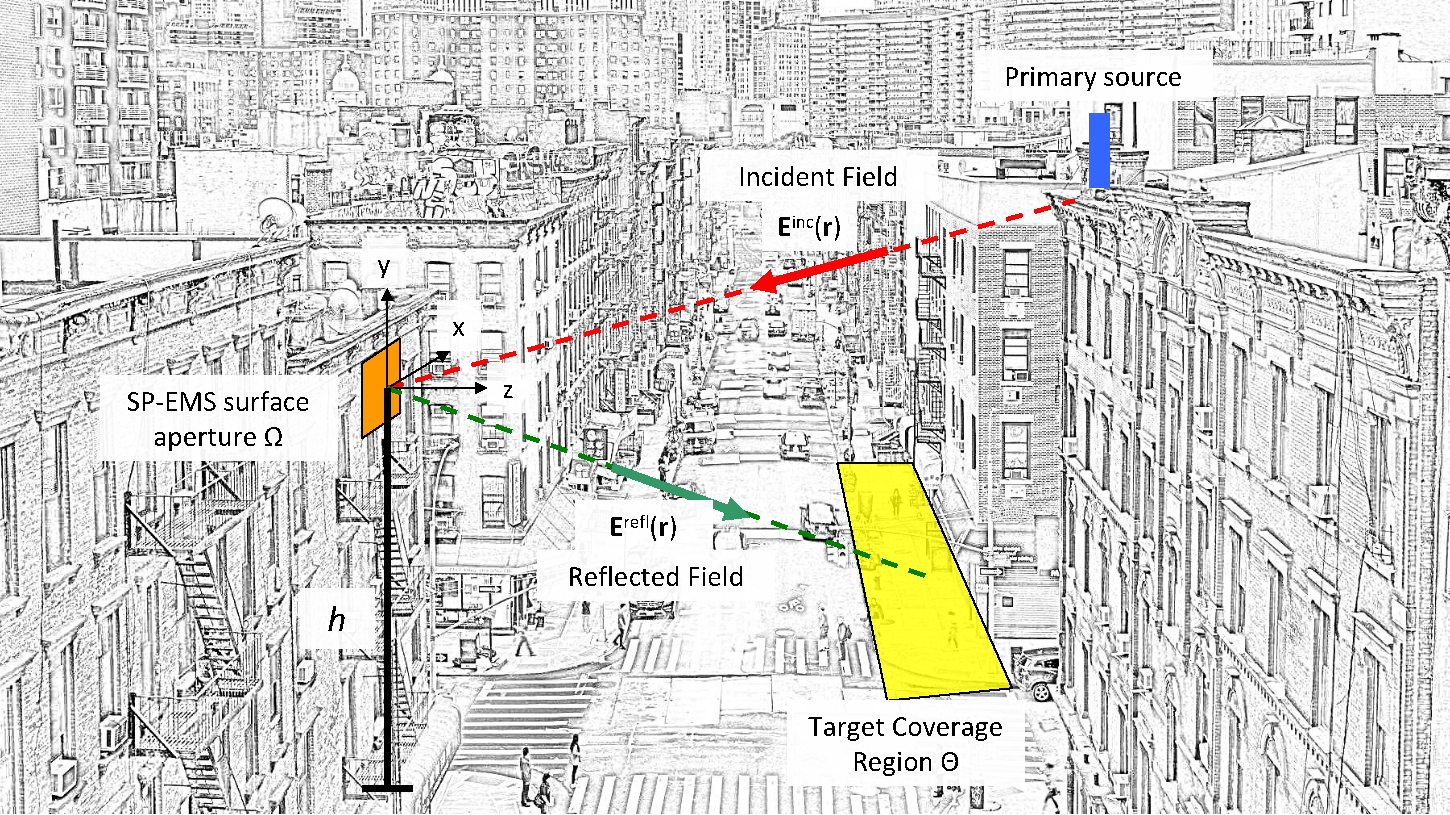}\tabularnewline
\tabularnewline
\tabularnewline
\end{tabular}\end{center}

\begin{center}~\vfill\end{center}

\begin{center}\textbf{Fig. 1 - G. Oliveri et} \textbf{\emph{al.}}\textbf{,}
{}``On the Improvement of the Performance of ...''\end{center}

\newpage
\begin{center}~\end{center}

\begin{center}\begin{tabular}{c}
\includegraphics[%
  width=0.40\columnwidth]{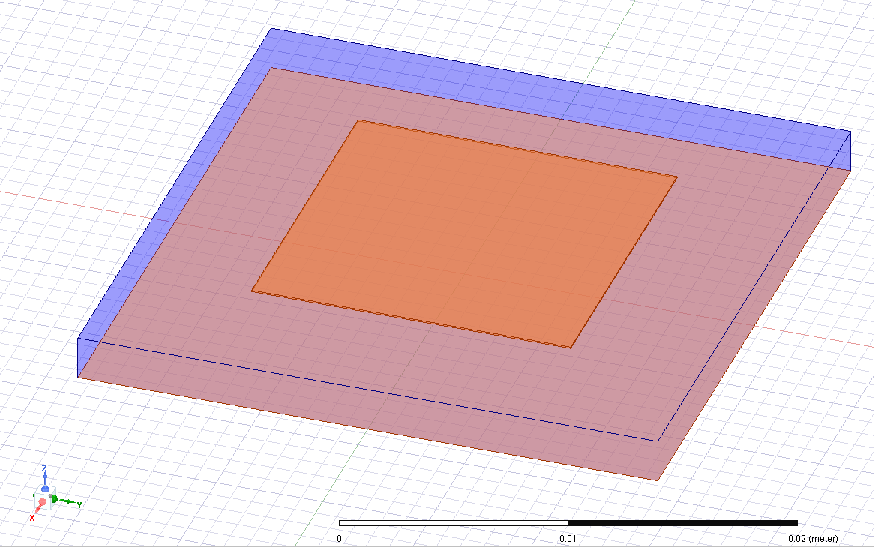}\tabularnewline
(\emph{a})\tabularnewline
\includegraphics[%
  width=0.70\columnwidth]{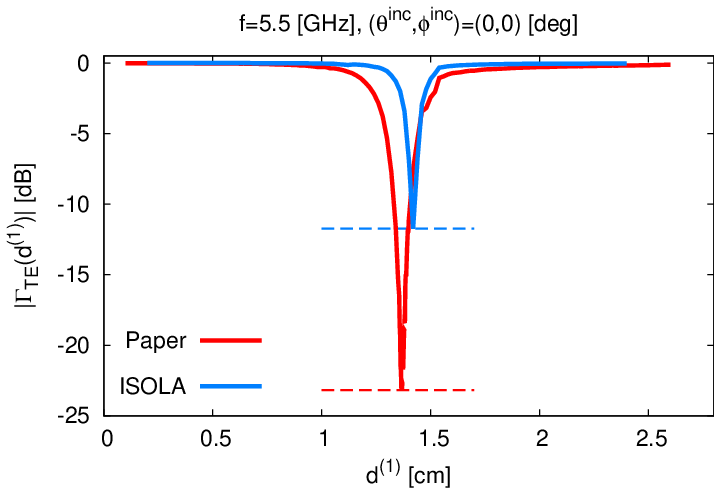}\tabularnewline
(\emph{b})\tabularnewline
\includegraphics[%
  width=0.70\columnwidth]{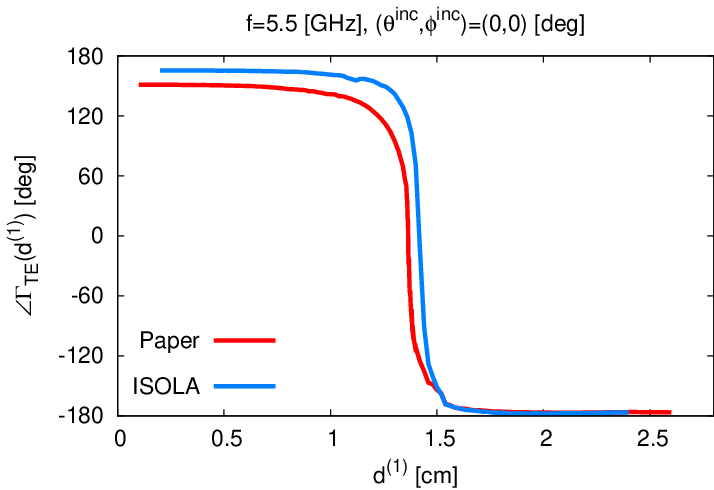}\tabularnewline
(\emph{c})\tabularnewline
\end{tabular}\end{center}

\begin{center}\textbf{Fig. 2 - G. Oliveri et} \textbf{\emph{al.}}\textbf{,}
{}``On the Improvement of the Performance of ...''\end{center}

\newpage
\begin{center}~\vfill\end{center}

\begin{center}\begin{tabular}{c}
\includegraphics[%
  width=0.90\columnwidth]{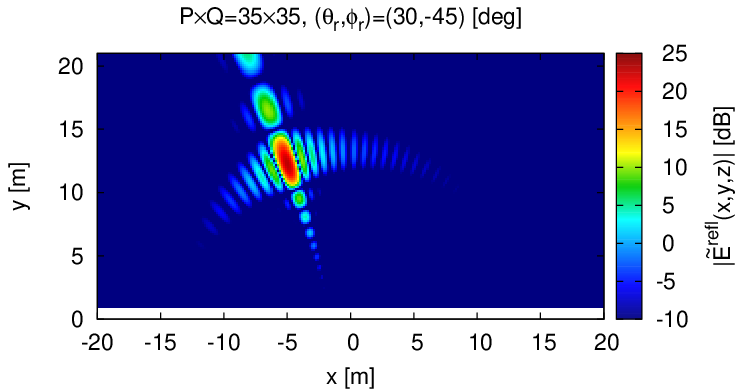}\tabularnewline
(\emph{a})\tabularnewline
\includegraphics[%
  width=0.80\columnwidth]{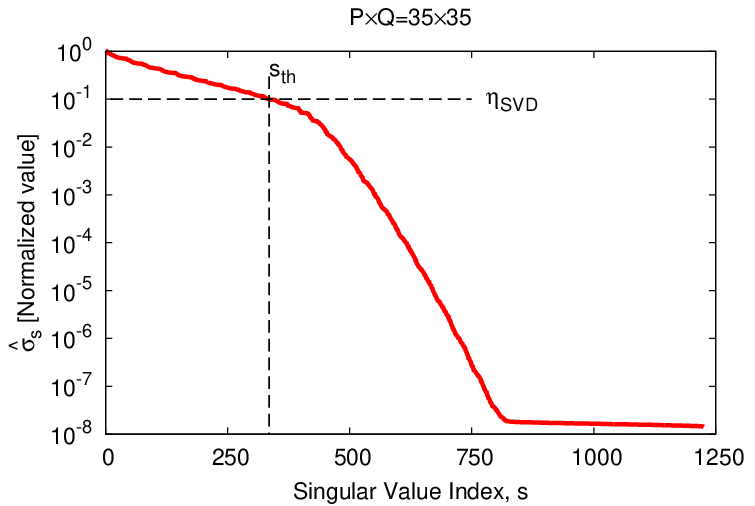}\tabularnewline
(\emph{b})\tabularnewline
\end{tabular}\end{center}

\begin{center}~\vfill\end{center}

\begin{center}\textbf{Fig. 3 - G. Oliveri et} \textbf{\emph{al.}}\textbf{,}
{}``On the Improvement of the Performance of ...''\end{center}

\newpage
\begin{center}~\vfill\end{center}

\begin{center}\begin{tabular}{c}
\includegraphics[%
  width=0.95\columnwidth]{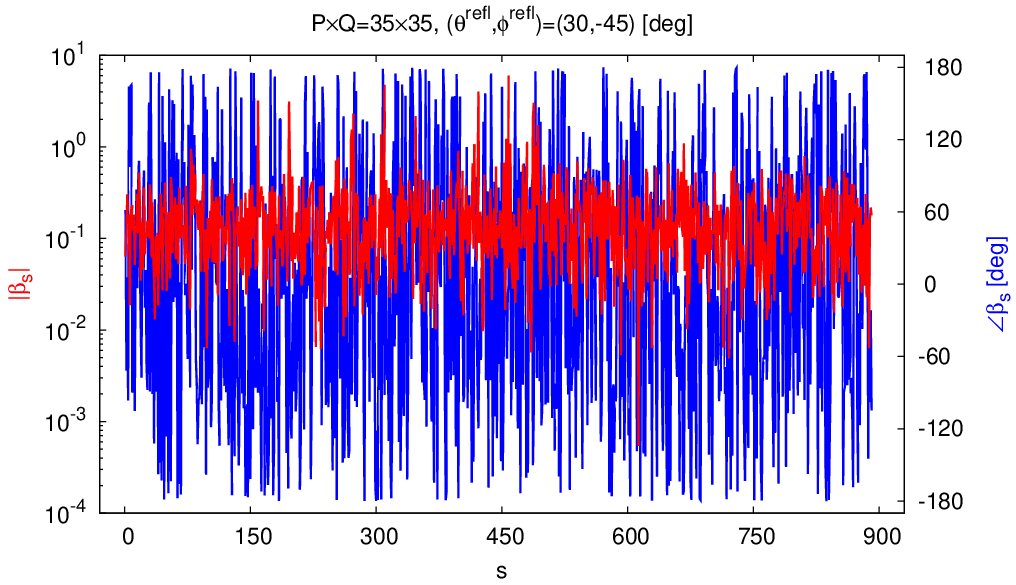}\tabularnewline
(\emph{a})\tabularnewline
\includegraphics[%
  width=0.55\columnwidth]{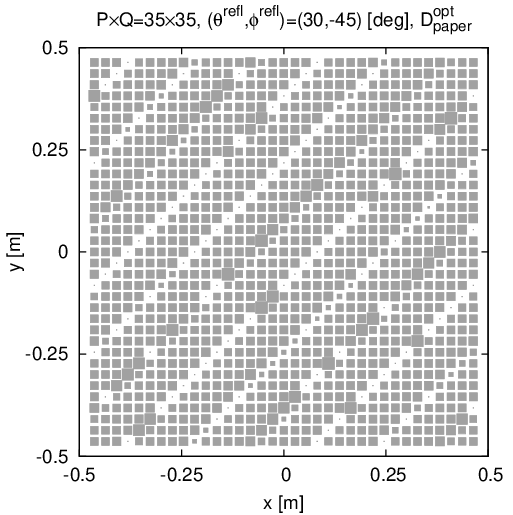}\tabularnewline
(\emph{b})\tabularnewline
\end{tabular}\end{center}

\begin{center}~\vfill\end{center}

\begin{center}\textbf{Fig. 4 - G. Oliveri et} \textbf{\emph{al.}}\textbf{,}
{}``On the Improvement of the Performance of ...''\end{center}

\newpage
\begin{center}\begin{sideways}
\begin{tabular}{cc}
\includegraphics[%
  width=0.69\columnwidth]{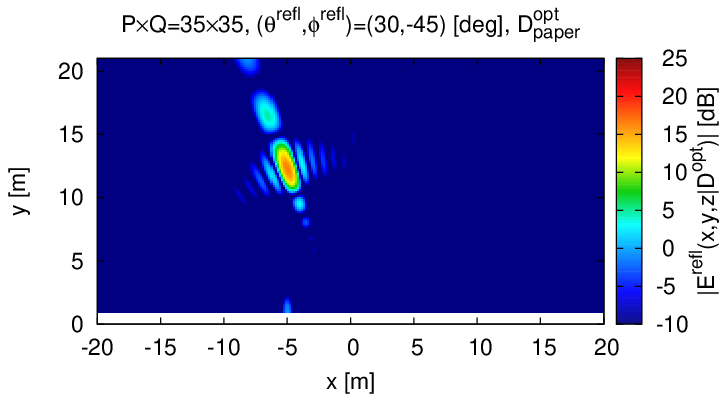}&
\includegraphics[%
  width=0.69\columnwidth]{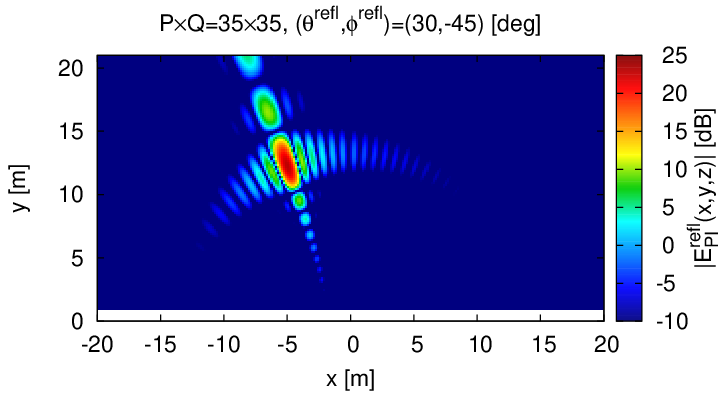}\tabularnewline
(\emph{a})&
(\emph{b})\tabularnewline
\includegraphics[%
  width=0.69\columnwidth]{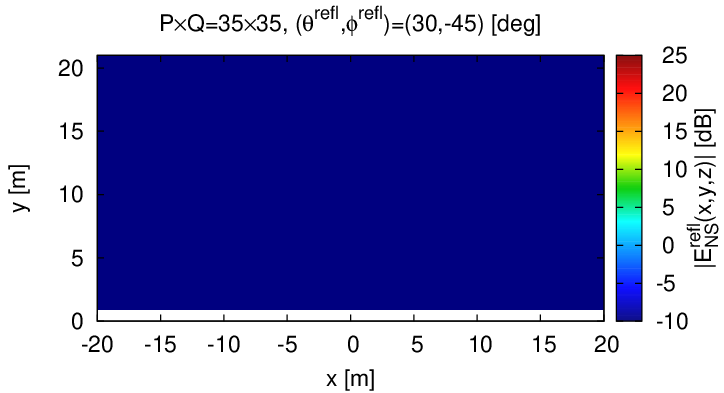}&
\includegraphics[%
  width=0.69\columnwidth]{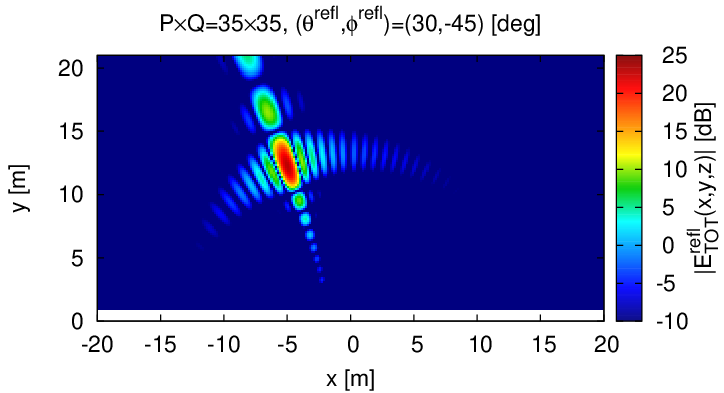}\tabularnewline
(\emph{c})&
(\emph{d})\tabularnewline
\end{tabular}
\end{sideways}\end{center}

\begin{center}\textbf{Fig. 5 - G. Oliveri et} \textbf{\emph{al.}}\textbf{,}
{}``On the Improvement of the Performance of ...''\end{center}

\newpage
\begin{center}~\vfill\end{center}

\begin{center}\begin{tabular}{cc}
\includegraphics[%
  width=0.45\columnwidth]{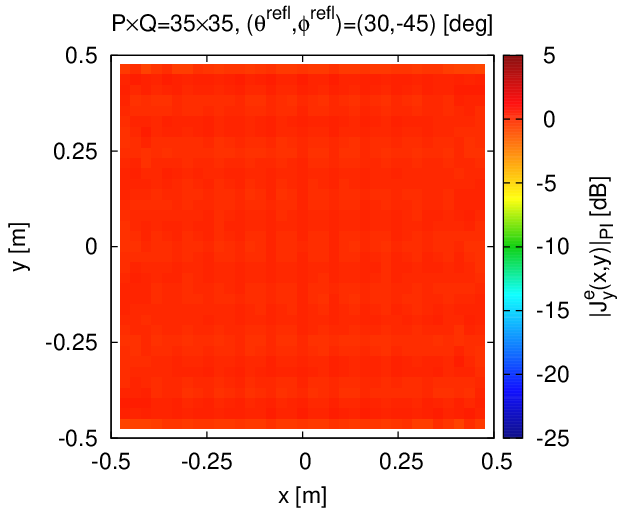}&
\includegraphics[%
  width=0.45\columnwidth]{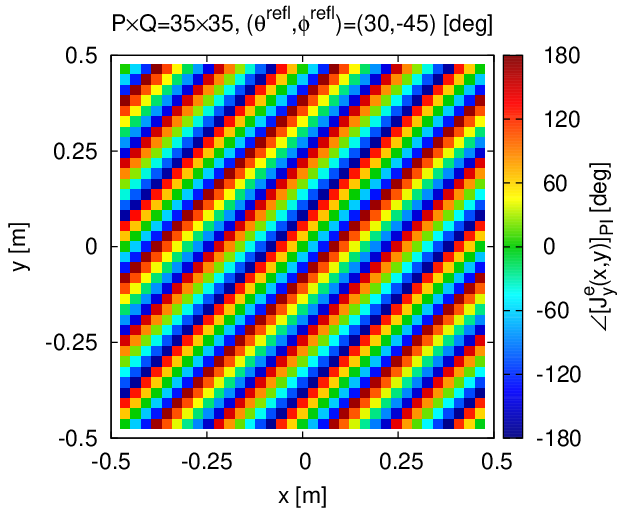}\tabularnewline
(\emph{a})&
(\emph{b})\tabularnewline
\includegraphics[%
  width=0.45\columnwidth]{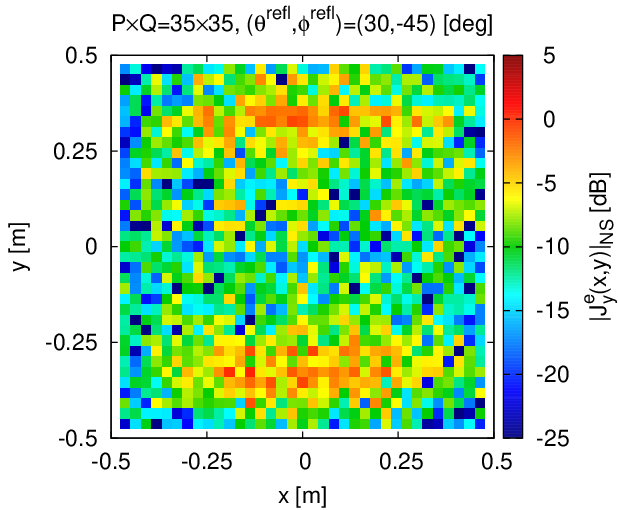}&
\includegraphics[%
  width=0.45\columnwidth]{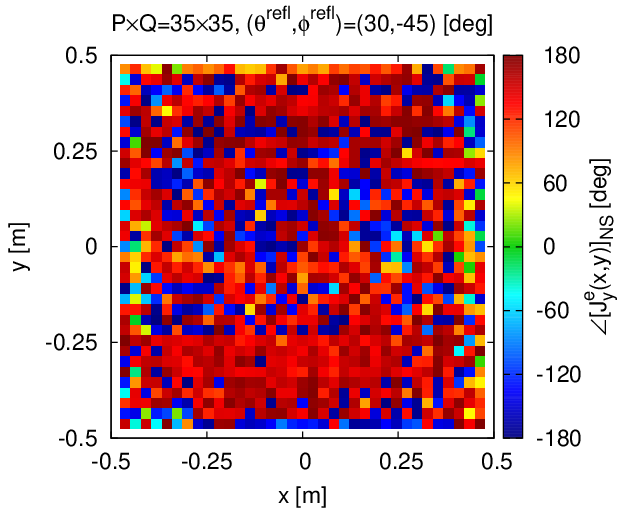}\tabularnewline
(\emph{c})&
(\emph{d})\tabularnewline
\includegraphics[%
  width=0.45\columnwidth]{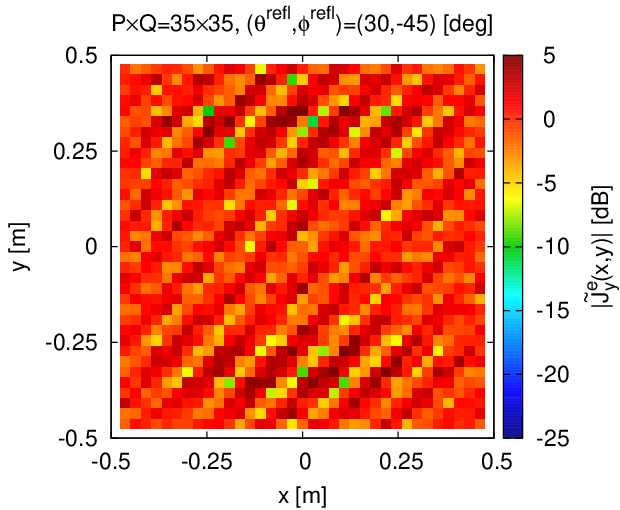}&
\includegraphics[%
  width=0.45\columnwidth]{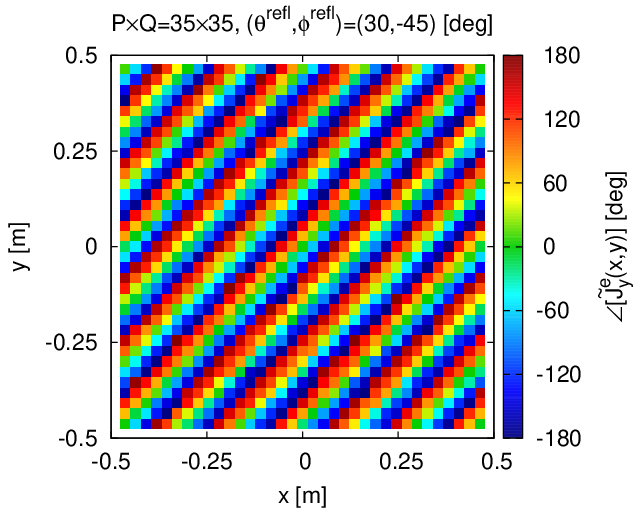}\tabularnewline
(\emph{e})&
(\emph{f})\tabularnewline
\end{tabular}\end{center}

\begin{center}~\vfill\end{center}

\begin{center}\textbf{Fig. 6 - G. Oliveri et} \textbf{\emph{al.}}\textbf{,}
{}``On the Improvement of the Performance of ...''\end{center}

\newpage
\begin{center}~\end{center}

\begin{center}\begin{tabular}{c}
\includegraphics[%
  width=0.40\columnwidth]{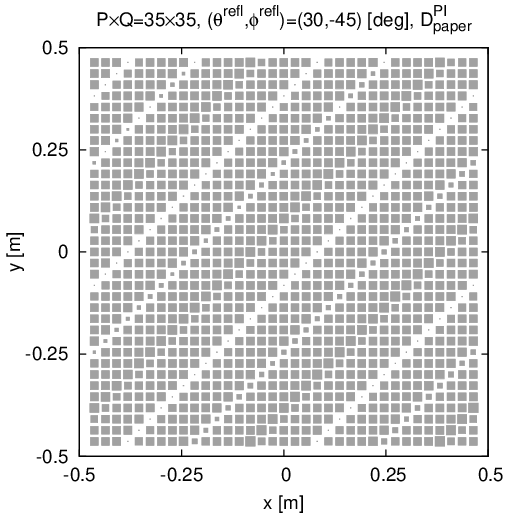}\tabularnewline
(\emph{a})\tabularnewline
\includegraphics[%
  width=0.73\columnwidth]{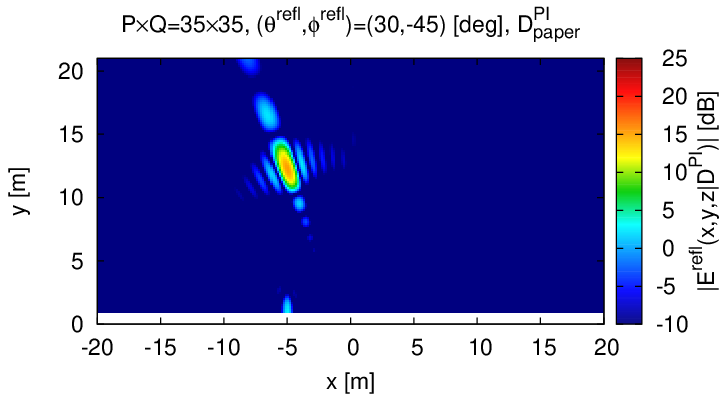}\tabularnewline
(\emph{b})\tabularnewline
\multicolumn{1}{c}{\includegraphics[%
  width=0.73\columnwidth]{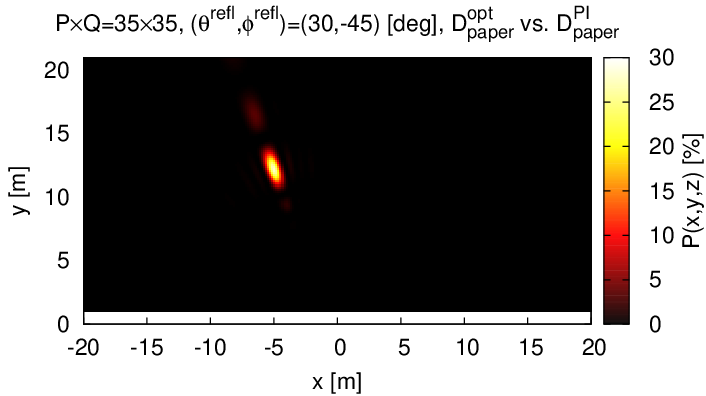}}\tabularnewline
\multicolumn{1}{c}{(\emph{c})}\tabularnewline
\end{tabular}\end{center}

\begin{center}~\end{center}

\begin{center}\textbf{Fig. 7 - G. Oliveri et} \textbf{\emph{al.}}\textbf{,}
{}``On the Improvement of the Performance of ...''\end{center}

\newpage
\begin{center}~\vfill\end{center}

\begin{center}\begin{tabular}{c}
\includegraphics[%
  width=0.70\columnwidth]{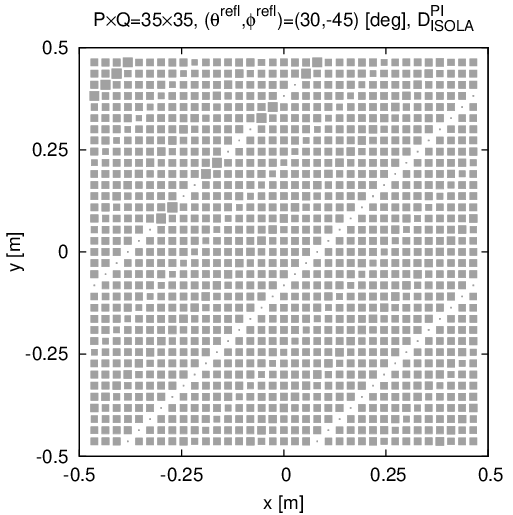}\tabularnewline
(\emph{a})\tabularnewline
\includegraphics[%
  width=0.90\columnwidth]{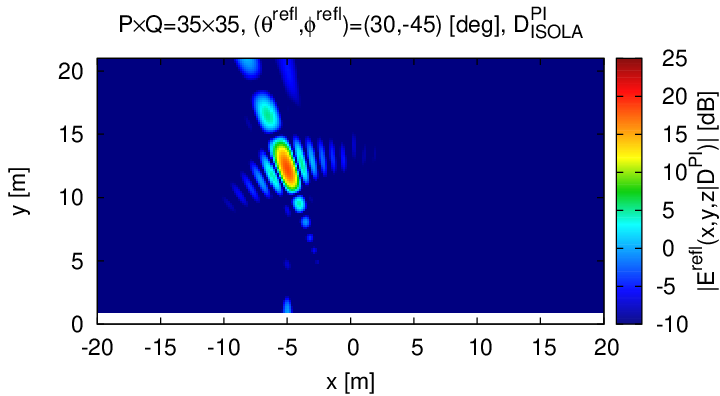}\tabularnewline
(\emph{b})\tabularnewline
\end{tabular}\end{center}

\begin{center}~\vfill\end{center}

\begin{center}\textbf{Fig. 8 - G. Oliveri et} \textbf{\emph{al.}}\textbf{,}
{}``On the Improvement of the Performance of ...''\end{center}

\newpage
\begin{center}~\end{center}

\begin{center}\begin{tabular}{c}
\includegraphics[%
  width=0.65\columnwidth]{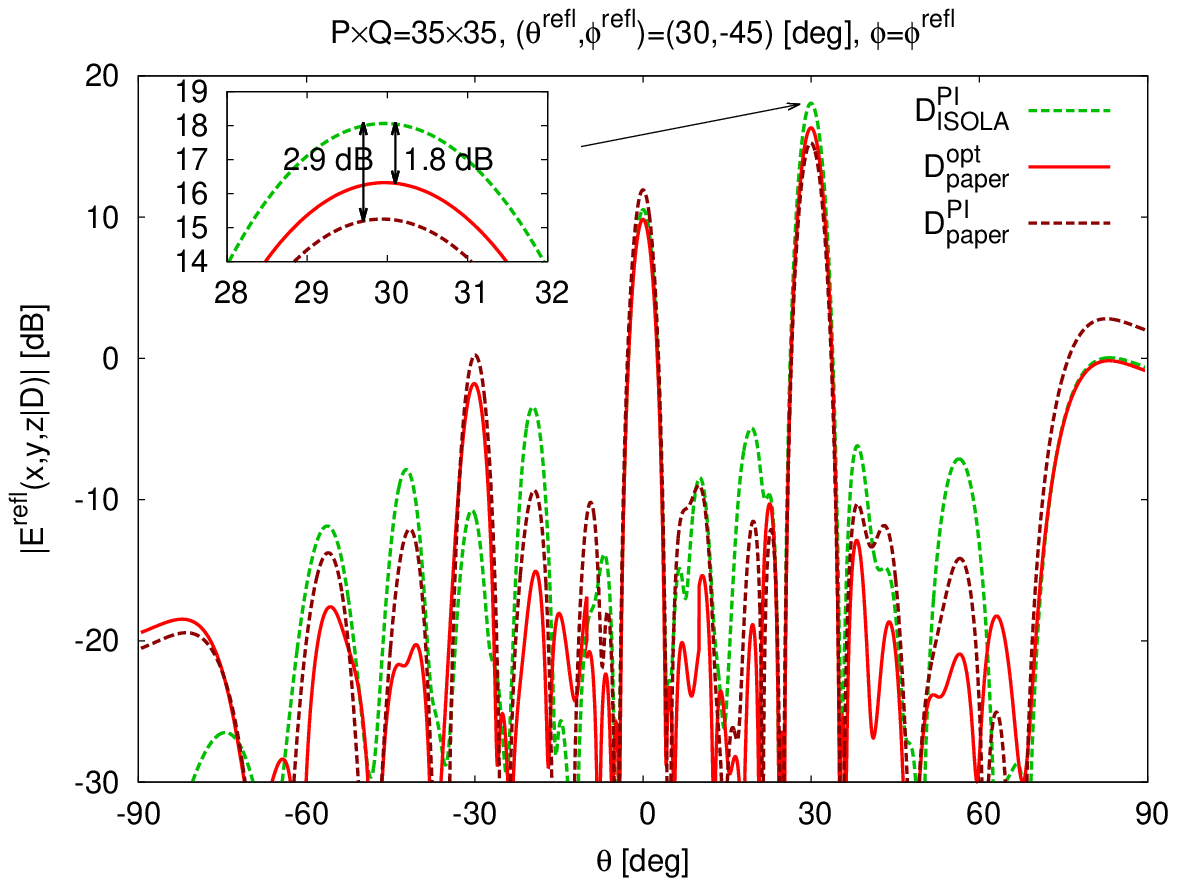}\tabularnewline
(\emph{a})\tabularnewline
\includegraphics[%
  width=0.65\columnwidth]{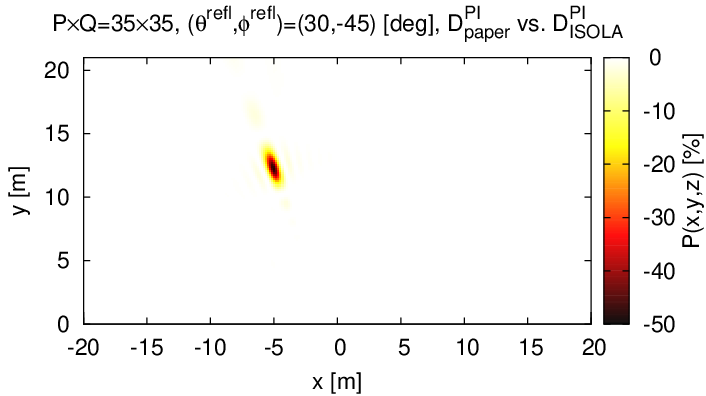}\tabularnewline
(\emph{b})\tabularnewline
\includegraphics[%
  width=0.65\columnwidth]{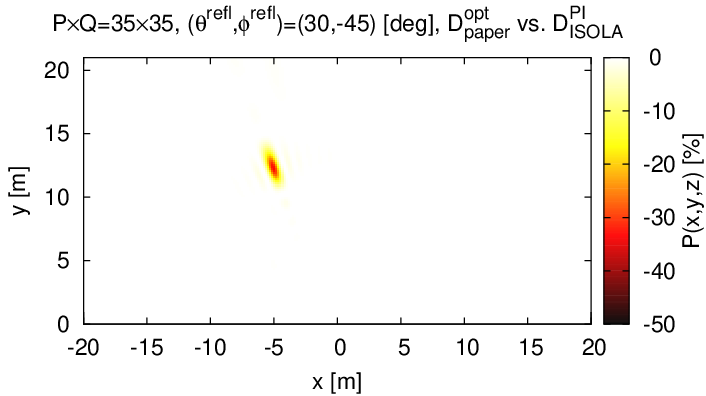}\tabularnewline
(\emph{c})\tabularnewline
\end{tabular}\end{center}

\begin{center}\textbf{Fig. 9 - G. Oliveri et} \textbf{\emph{al.}}\textbf{,}
{}``On the Improvement of the Performance of ...''\end{center}

\newpage
\begin{center}~\vfill\end{center}

\begin{center}\begin{tabular}{c}
\includegraphics[%
  width=0.85\columnwidth]{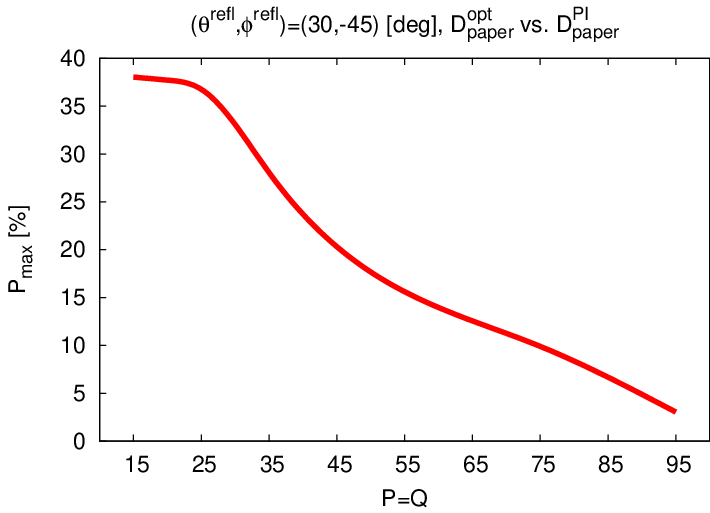}\tabularnewline
(\emph{a})\tabularnewline
\includegraphics[%
  width=0.85\columnwidth]{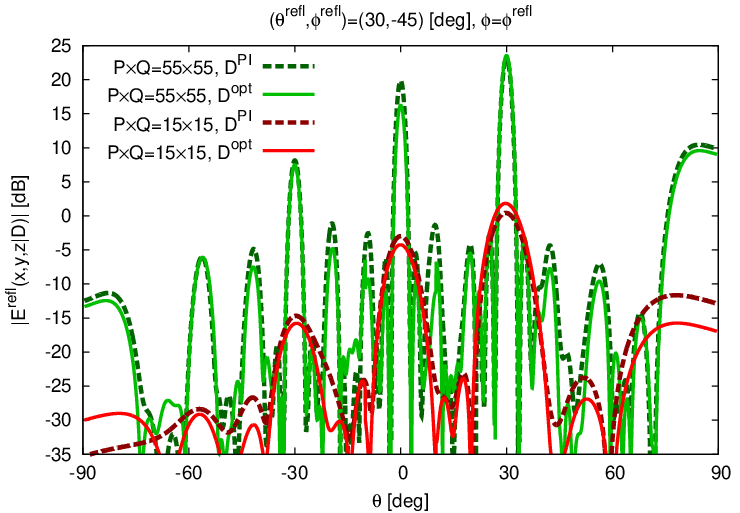}\tabularnewline
(\emph{b})\tabularnewline
\end{tabular}\end{center}

\begin{center}~\vfill\end{center}

\begin{center}\textbf{Fig. 10 - G. Oliveri et} \textbf{\emph{al.}}\textbf{,}
{}``On the Improvement of the Performance of ...''\end{center}

\newpage
\begin{center}~\vfill\end{center}

\begin{center}\includegraphics[%
  width=0.95\columnwidth]{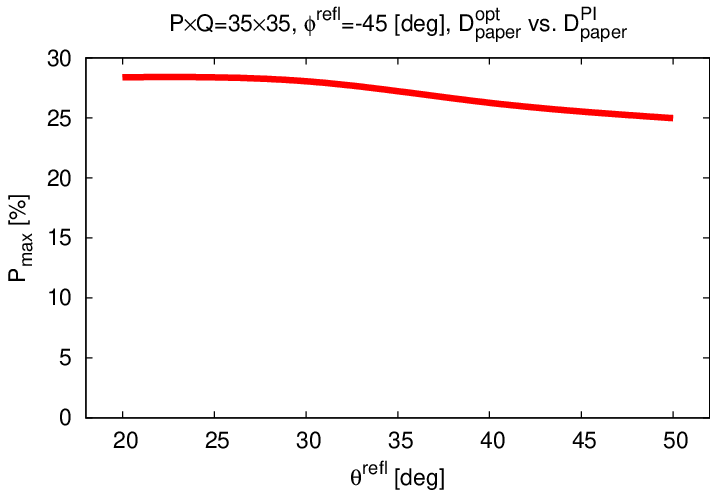}\end{center}

\begin{center}~\vfill\end{center}

\begin{center}\textbf{Fig. 11 - G. Oliveri et} \textbf{\emph{al.}}\textbf{,}
{}``On the Improvement of the Performance of ...''\end{center}

\newpage
\begin{center}~\vfill\end{center}

\begin{center}\begin{tabular}{c}
\includegraphics[%
  width=0.75\columnwidth]{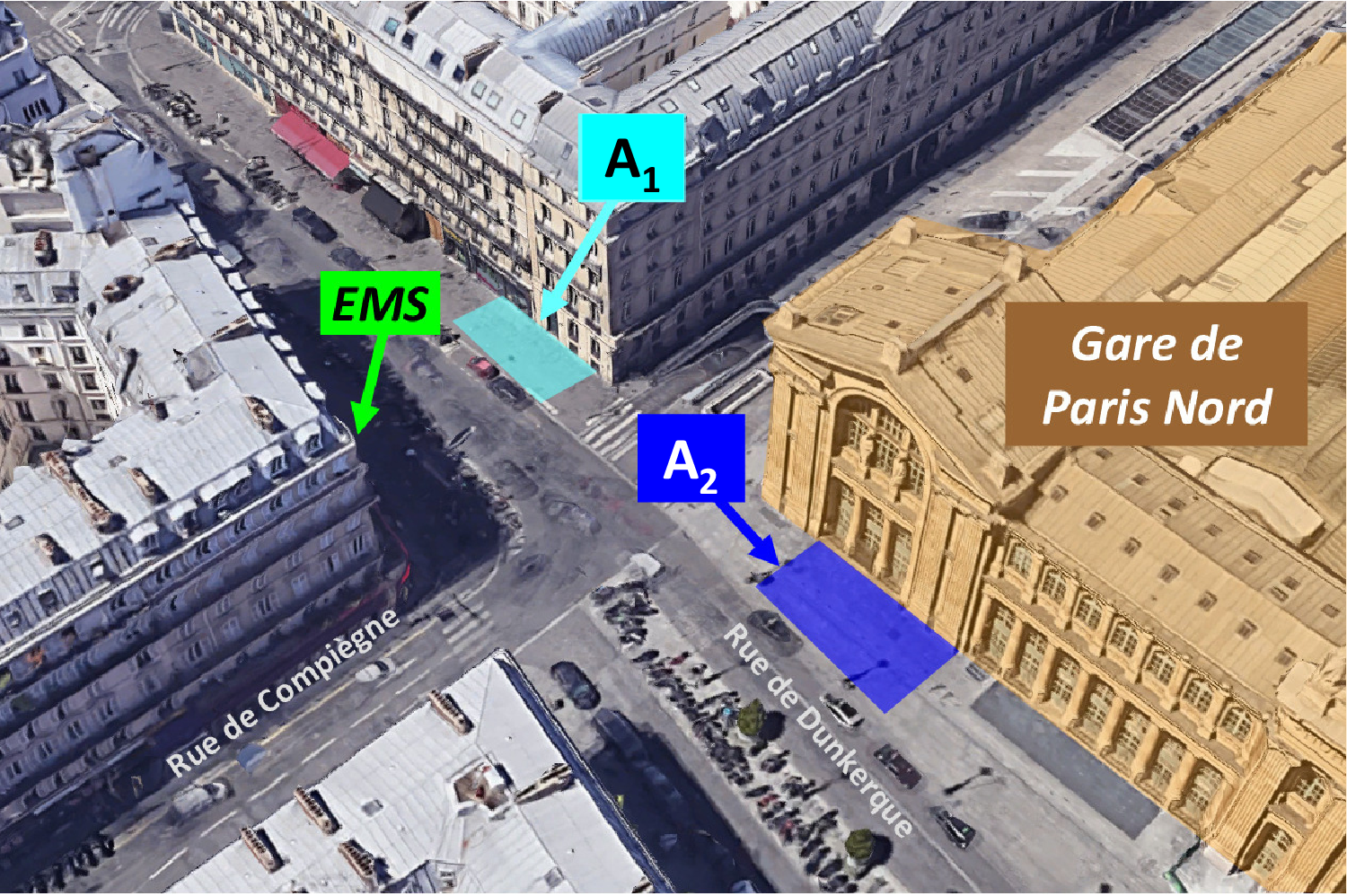}\tabularnewline
(\emph{a})\tabularnewline
\includegraphics[%
  width=0.90\columnwidth]{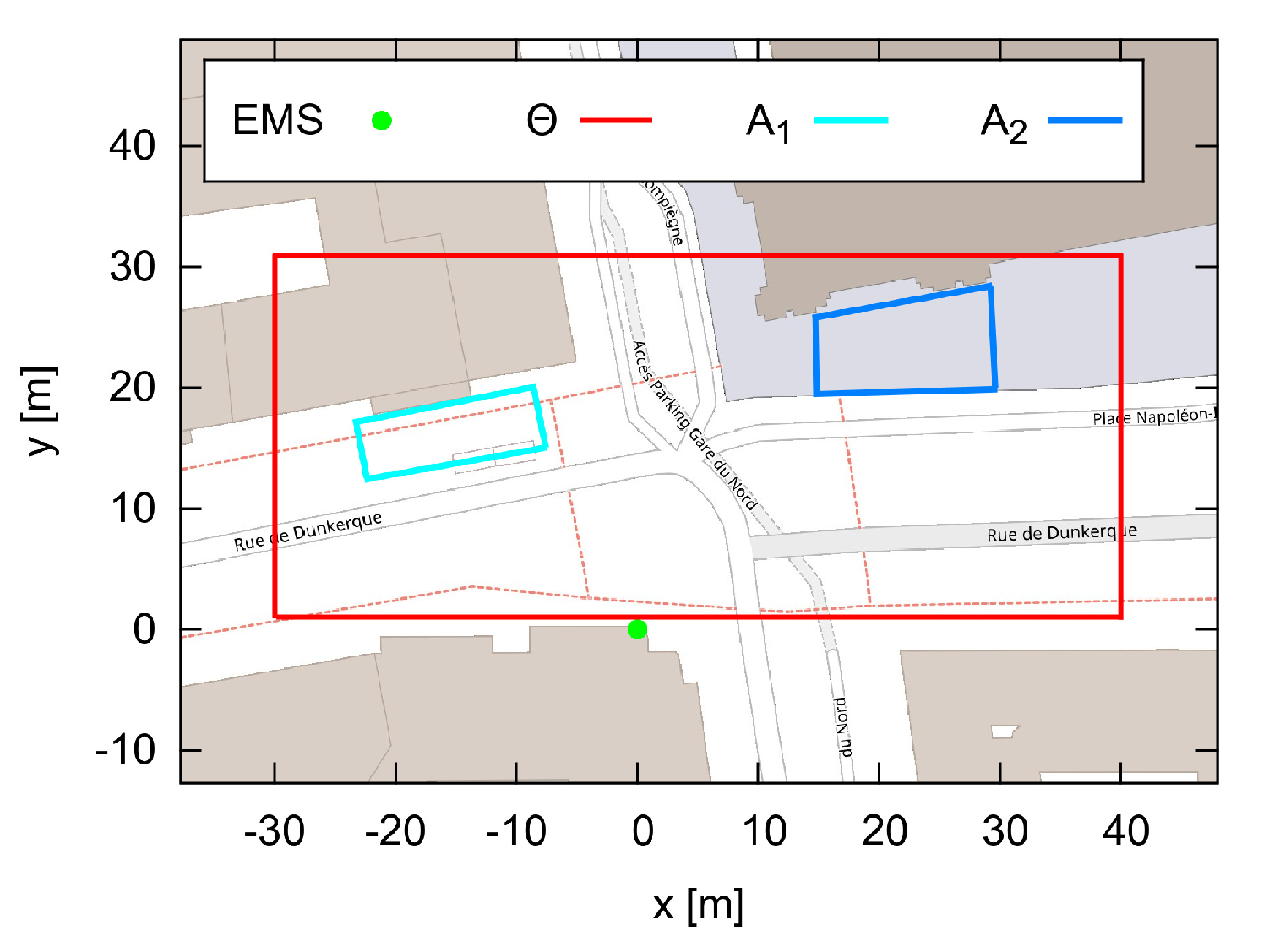}\tabularnewline
(\emph{b})\tabularnewline
\end{tabular}\end{center}

\begin{center}~\vfill\end{center}

\begin{center}\textbf{Fig. 12 - G. Oliveri et} \textbf{\emph{al.}}\textbf{,}
{}``On the Improvement of the Performance of ...''\end{center}

\newpage
\begin{center}~\vfill\end{center}

\begin{center}\begin{tabular}{c}
\includegraphics[%
  width=0.90\columnwidth]{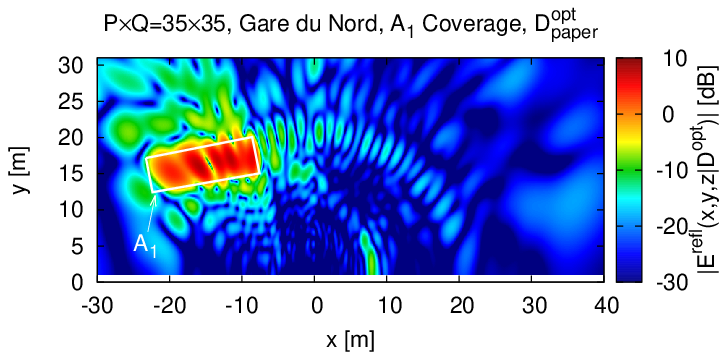}\tabularnewline
(\emph{a})\tabularnewline
\includegraphics[%
  width=0.90\columnwidth]{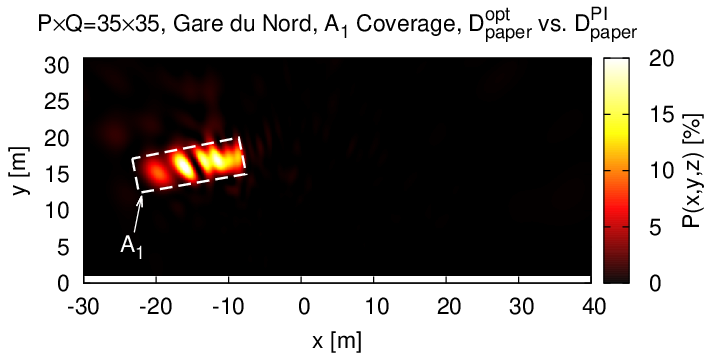}\tabularnewline
(\emph{b})\tabularnewline
\end{tabular}\end{center}

\begin{center}~\vfill\end{center}

\begin{center}\textbf{Fig. 13 - G. Oliveri et} \textbf{\emph{al.}}\textbf{,}
{}``On the Improvement of the Performance of ...''\end{center}

\newpage
\begin{center}~\vfill\end{center}

\begin{center}\begin{tabular}{c}
\includegraphics[%
  width=0.90\columnwidth]{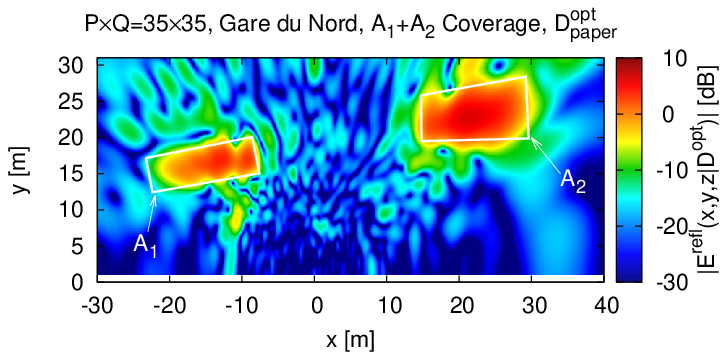}\tabularnewline
(\emph{a})\tabularnewline
\includegraphics[%
  width=0.90\columnwidth]{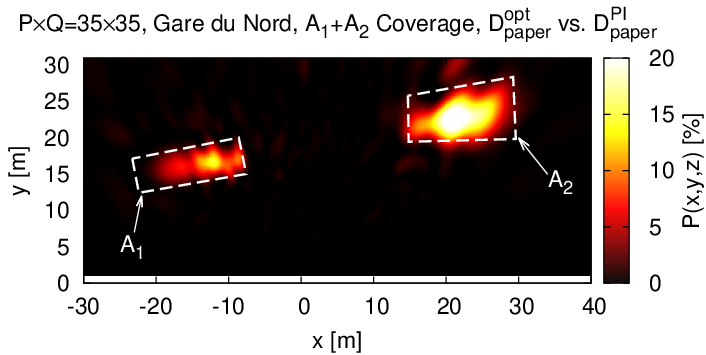}\tabularnewline
(\emph{b})\tabularnewline
\end{tabular}\end{center}

\begin{center}~\vfill\end{center}

\begin{center}\textbf{Fig. 14 - G. Oliveri et} \textbf{\emph{al.}}\textbf{,}
{}``On the Improvement of the Performance of ...''\end{center}

\newpage
\begin{center}~\vfill\end{center}

\begin{center}\begin{tabular}{cc}
\includegraphics[%
  width=0.28\columnwidth,
  keepaspectratio]{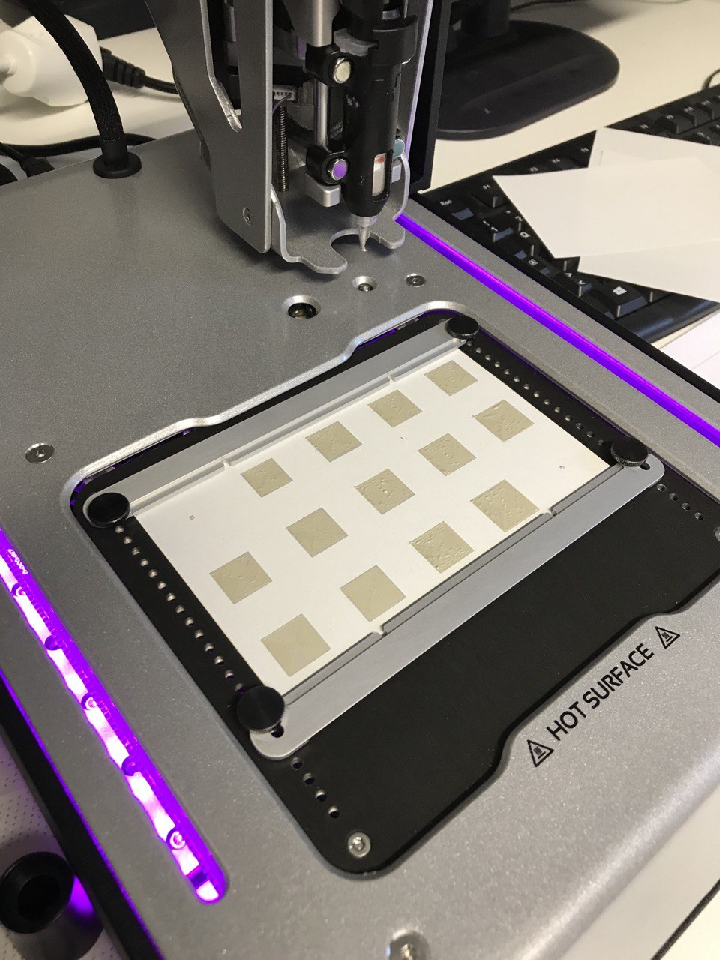}&
\includegraphics[%
  width=0.50\columnwidth]{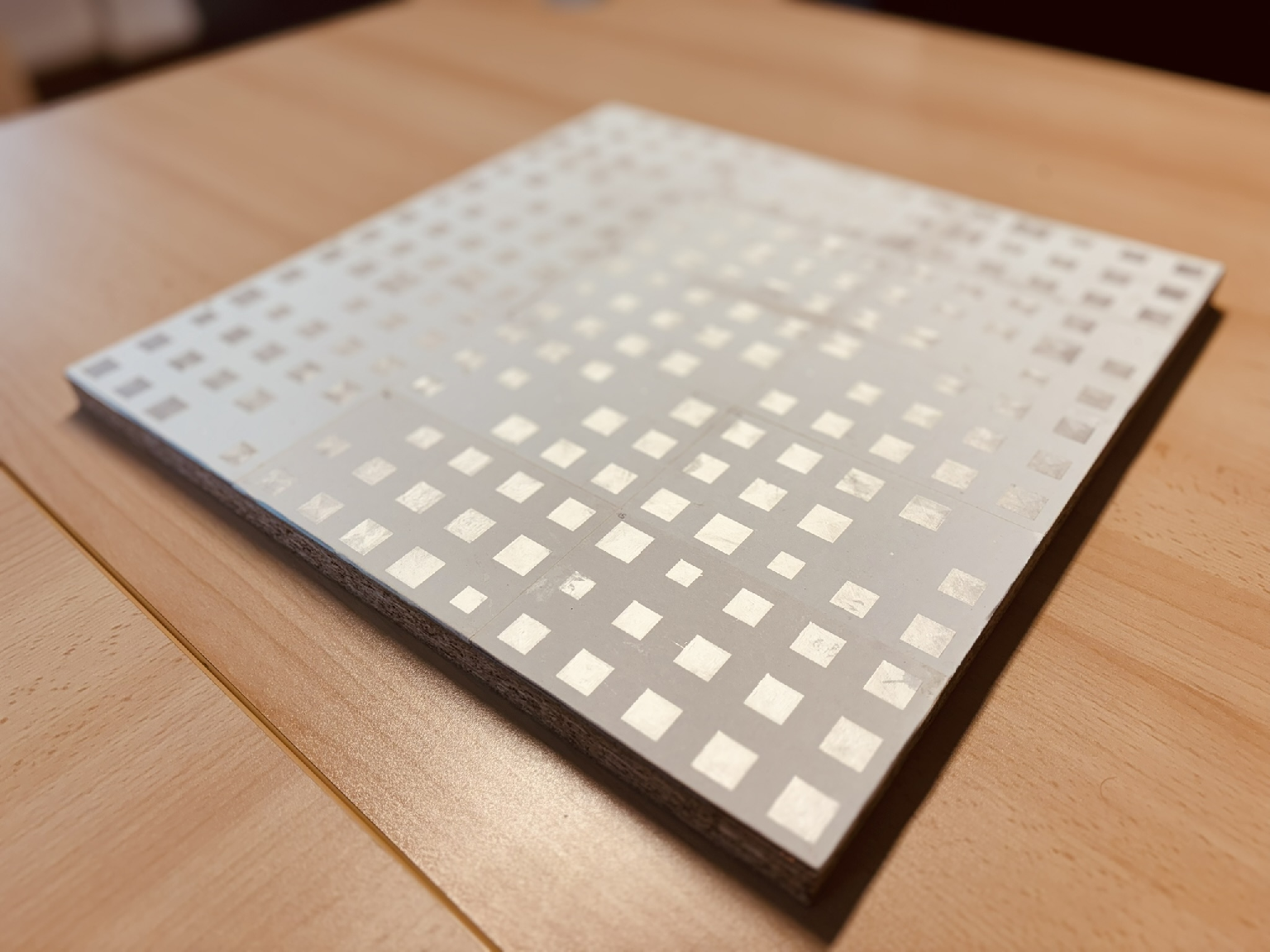}\tabularnewline
(\emph{a})&
(\emph{b})\tabularnewline
\multicolumn{2}{c}{\includegraphics[%
  width=0.95\columnwidth,
  keepaspectratio]{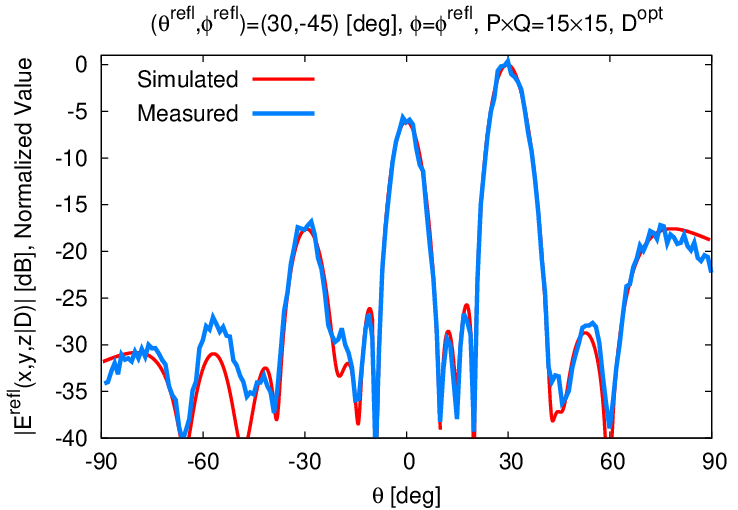}}\tabularnewline
\multicolumn{2}{c}{(\emph{c})}\tabularnewline
\end{tabular}\end{center}

\begin{center}~\vfill\end{center}

\begin{center}\textbf{Fig. 15 - G. Oliveri et} \textbf{\emph{al.}}\textbf{,}
{}``On the Improvement of the Performance of ...''\end{center}

\begin{thebibliography}{10}
\bibitem{Massa 2021}A. Massa, A. Benoni, P. Da Ru, S. K. Goudos, B. Li, G. Oliveri, A.
Polo, P. Rocca, and M. Salucci, {}``Designing smart electromagnetic
environments for next-generation wireless communications,'' \emph{Telecom},
vol. 2, no. 2, pp. 213-221, 2021.
\bibitem{Yang 2022}F. Yang, D. Erricolo, and A. Massa, {}``Guest Editorial - Smart Electromagnetic
Environment,'' \emph{IEEE Trans. Antennas Propag}., vol. 70, no.
10, pp. 8687-8690, Oct. 2022.
\bibitem{Barbuto 2022}M. Barbuto, Z. Hamzavi-Zarghani, M. Longhi, A. Monti, D. Ramaccia,
S. Vellucci, A. Toscano, and F. Bilotti, {}``Metasurfaces 3.0: a
new paradigm for enabling smart electromagnetic environments,'' \emph{IEEE
Trans. Antennas Propag}., vol. 70, no. 10, pp. 8883-8897, Oct. 2022.
\bibitem{Di Renzo 2019}M. Di Renzo, M. Debbah, D.-T. Phan-Huy, A. Zappone, M.-S. Alouini,
C. Yuen, V. Sciancalepore, G. C. Alexandropoulos, J. Hoydis, H. Gacanin,
J. De Rosny, A. Bounceur, G. Lerosey, and M. Fink, {}``Smart radio
environments empowered by reconfigurable AI meta-surfaces: An idea
whose time has come,'' \emph{EURASIP J. Wireless Commun. Netw.},
vol. 129, pp. 1-20, 2019.
\bibitem{Di Renzo 2020}M. Di Renzo, A. Zappone, M. Debbah, M.-S. Alouini, C. Yuen, J. De
Rosny, and S. Tretyakov, {}``Smart radio environments empowered by
reconfigurable intelligent surfaces: How it works, state of research,
and the road ahead,'' \emph{IEEE J. Sel. Areas Comm}., vol. 38, no.
11, pp. 2450-2525, Nov. 2020.
\bibitem{Di Renzo 2020b}M. Di Renzo, K. Ntontin, J. Song, F. H. Danufane, X. Qian, F. Lazarakis,
J. De Rosny, D.-T. Phan-Huy, O. Simeone, R. Zhang, M. Debbah, G. Lerosey,
M. Fink, S. Tretyakov, and S. Shamai, {}``Reconfigurable intelligent
surfaces vs. relaying: Differences, similarities, and performance
comparison,'' \emph{IEEE Open J. Comm. Soc}., vol. 1, pp. 798-807,
2020.
\bibitem{Huang 2019}C. Huang, A. Zappone, G. C. Alexandropoulos, M. Debbah, and C. Yuen,
{}``Reconfigurable intelligent surfaces for energy efficiency in
wireless communication,'' \emph{IEEE Trans. Wireless Commun.}, vol.
18, no. 8, pp. 4157-4170, Aug. 2019.
\bibitem{Oliveri 2021c}G. Oliveri, P. Rocca, M. Salucci, and A. Massa, {}``Holographic smart
EM skins for advanced beam power shaping in next generation wireless
environments,'' \emph{IEEE J. Multiscale Multiphysics Computat. Techn.},
vol. 6, pp. 171-182, Oct. 2021.
\bibitem{Oliveri 2022}G. Oliveri, F. Zardi, P. Rocca, M. Salucci, and A. Massa, {}``Building
a smart EM environment - AI-Enhanced aperiodic micro-scale design
of passive EM skins,'' \emph{IEEE Trans. Antennas Propag.}, vol.
70, no. 10, pp. 8757-8770, Oct. 2022.
\bibitem{Oliveri 2023}G. Oliveri, F. Zardi, P. Rocca, M. Salucci and A. Massa, {}``Constrained
design of passive static EM skins,'' \emph{IEEE Trans. Antennas Propag.},
vol. 71, no. 2, pp. 1528-1538, Feb. 2023.
\bibitem{Oliveri 2023b}G. Oliveri, M. Salucci, and A. Massa, {}``Generalized Analysis and
Unified Design of EM Skins,'' \emph{IEEE Trans. Antennas Propag.},
vol. 71, no.8, pp. 6579-6592, Aug.2023.
\bibitem{Vaquero 2024}A. F. Vaquero, E. Martinez-de-Rioja, M. Arrebola, J. A. Encinar and
M. Achour, {}``Smart electromagnetic skin to enhance near-field coverage
in mm-Wave 5G indoor scenarios,'' \emph{IEEE Trans. Antennas Propag.},
vol. 72, no. 5, pp. 4311-4326, May 2024.
\bibitem{Yang 2019}F. Yang and Y. Rahmat-Samii, \emph{Surface Electromagnetics with Applications
in Antenna, Microwave, and Optical Engineering}, Cambridge, UK: Cambridge
University Press, 2019.
\bibitem{Alu 2024}A. Alu, N. Engheta, A. Massa, and G. Oliveri, Eds., \emph{Metamaterials-by-Design:
Theory, Technologies, and Vision}. Amsterdam, NL: Elsevier, 2024.
\bibitem{Rocca 2009}P. Rocca, M. Benedetti, M. Donelli, D. Franceschini, and A. Massa,
{}``Evolutionary optimization as applied to inverse problems,''
\emph{Inv. Probl}., vol. 25, art no. 123003, pp. 1-41, Dec. 2009.
\bibitem{Yang 2007a}L. Yang, A. Rida, R. Vyas, and M. M. Tentzeris, {}``RFID tag and
RF structures on a paper substrate using inkjet-printing technology,''
\emph{IEEE Trans. Microw. Theory Techn.}, vol. 55, no. 12, pp. 2894-2901,
Dec. 2007.
\bibitem{Lindell 2019}I. V. Lindell and A. Sihvola, \emph{Boundary Conditions in Electromagnetics}.
IEEE Press, 2019.
\bibitem{Osipov 2017}A. Osipov and S. Tretyakov, \emph{Modern electromagnetic scattering
theory with applications.} John Wiley \& Sons, 2017.
\bibitem{Salucci 2018c}M. Salucci, L. Tenuti, G. Oliveri, and A. Massa, {}``Efficient prediction
of the EM response of reflectarray antenna elements by an advanced
statistical learning method,'' \emph{IEEE Trans. Antennas Propag}.,
vol. 66, no. 8, pp. 3995-4007, Aug. 2018.
\bibitem{Oliveri 2022b}G. Oliveri, M. Salucci, and A. Massa, {}``Towards efficient reflectarray
digital twins - An EM-driven machine learning perspective,'' \emph{IEEE
Trans. Antennas Propag.}, vol. 70, no. 7, pp. 5078-5093, Jul. 2022.
\bibitem{HFSS 2021}ANSYS Electromagnetics Suite - HFSS (2021). ANSYS, Inc.
\bibitem{Bertero 1998}M. Bertero and P. Boccacci, \emph{Introduction to Inverse Problems
in Imaging}, U.K., Bristol: IoP Publishing, 1998.
\bibitem{van den Berg 2002}P. M. van den Berg and A. Abubakar, {}``Inverse scattering algorithms
based on contrast source integral representations,'' \emph{Inverse
Probl. Eng.}, vol. 10, no. 6, pp. 559-576, 2002.
\bibitem{Kawahara 2014}Y. Kawahara, S. Hodges, N.-W. Gong, S. Olberding, and J. Steimle,
{}``Building functional prototypes using conductive inkjet printing,''
\emph{IEEE Pervasive Comput.}, vol. 13, no. 3, pp. 30-38, Jul. 2014.
\bibitem{Alimenti 2012}F. Alimenti, P. Mezzanotte, M. Dionigi, M. Virili, and L. Roselli,
{}``Microwave circuits in paper substrates exploiting conductive
adhesive tapes,'' \emph{IEEE Microw. Wireless Compon. Lett.}, vol.
22, no. 12, pp. 660-662, Dec. 2012.
\bibitem{Borgese 2017}M. Borgese, F. A. Dicandia, F. Costa, S. Genovesi, and G. Manara,
{}``An inkjet printed chipless RFID sensor for wireless humidity
monitoring,'' \emph{IEEE Sensors J.}, vol. 17, no. 15, pp. 4699-4707,
Aug. 2017.
\bibitem{Genovesi 2016}S. Genovesi, F. Costa, F. Fanciulli, and A. Monorchio, {}``Wearable
inkjet-printed wideband antenna by using miniaturized AMC for sub-GHz
applications,'' \emph{IEEE Antennas Wireless Propag. Lett.}, vol.
15, pp. 1927-1930, Jan. 2016.
\bibitem{Legay 2013}H. Legay, D. Bresciani, E. Labiole, R. Chiniard, and R. Gillard, {}``A
multi facets composite panel reflectarray antenna for a space contoured
beam antenna in Ku band,'' \emph{PIER B}, vol. 54, pp. 1-26, 2013.
\bibitem{Rocca 2022}P. Rocca, P. Da Ru, N. Anselmi, M. Salucci, G. Oliveri, D. Erricolo,
and A. Massa, {}``On the design of modular reflecting EM skins for
enhanced urban wireless coverage,'' \emph{IEEE Trans. Antennas Propag}.,
vol. 70, no. 10, pp. 8771-8784, Oct. 2022.
\end{thebibliography}
\end{document}